\RequirePackage{fix-cm}
\documentclass[twocolumn]{svjour3}           
\usepackage[utf8]{inputenc}
\usepackage{amsmath,graphicx,makecell}
\usepackage{subfigure}
\usepackage{enumitem}
\usepackage{booktabs}
\usepackage{multirow}
\usepackage{mathptmx}      
\usepackage{adjustbox}
\usepackage{graphicx}
\usepackage{amssymb}
\usepackage{epstopdf}

\newlength\savedwidth

\newlength\savewidth
\newcommand\shline{\noalign{\global\savewidth\arrayrulewidth
                            \global\arrayrulewidth 1.0pt}%
                   \hline
                   \noalign{\global\arrayrulewidth\savewidth}}
\begin{document}
\title{Hybrid integration of multilayer perceptrons and parametric models for reliability forecasting in the smart grid}
\author{Longfei {WEI}$^{1}$, Arif I. {SARWAT}$^{1}$}


\date{Received: date / Accepted: date}

\institute{Arif I. {SARWAT} \at
\email{asarwat@fiu.edu} \\
\and
Longfei {WEI} \at
            \email{lwei004@fiu.edu}           \\
\at $^{1}$ {Department of Electrical and Computer Engineering, Florida International University, Miami, FL 33174, USA}
}
\maketitle

\begin{abstract}
The reliable power system operation is a major goal for electric utilities, which requires the accurate reliability forecasting to minimize the duration of power interruptions. Since weather conditions are usually the leading causes for power interruptions in the smart grid, especially for its distribution networks, this paper comprehensively investigates the combined effect of various weather parameters on the reliability performance of distribution networks. Specially, a multilayer perceptron (MLP) based framework is proposed to forecast the daily numbers of sustained and momentary power interruptions in one distribution management area using time series of common weather data. First, the parametric regression models are implemented to analyze the relationship between the daily numbers of power interruptions and various common weather parameters, such as temperature, precipitation, air pressure, wind speed, and lightning. The selected weather parameters and corresponding parametric models are then integrated as inputs to formulate a MLP neural network model to predict the daily numbers of power interruptions. A modified extreme learning machine (ELM) based hierarchical learning algorithm is introduced for training the formulated model using real-time reliability data from an electric utility in Florida and common weather data from National Climatic Data Center (NCDC). In addition, the sensitivity analysis is implemented to determine the various impacts of different weather parameters on the daily numbers of power interruptions.
\keywords{MLP neural network, Power interruption, Reliability forecasting, Smart grid, Weather condition}
\end{abstract}

\section{Introduction} \label{intro}

The reliability has always been a critical focus area for the design and operation of the electric power grid, where the distribution networks account for up to $90$\% of all customer reliability problems~\cite{Re0,Re1}. Improving the distribution reliability is a key point for increasing customer satisfaction and system performance. However, with the high penetration of renewable energy resources and increasing electricity demands in distribution networks, meeting reliability objectives in modern grids becomes increasingly challenging~\cite{Con1,Re11,Re12,Re13}. According to the Lawrence Berkeley National Laboratory's (L\-BNL) report in 2016~\cite{Re2,Re21}, the annual cost for power interruptions to the electricity customers of the United States is estimated to be \$110 billions, which increases more than 30\% since 2004.

\begin{figure}[b!]
\centering
  \includegraphics[width=8cm]{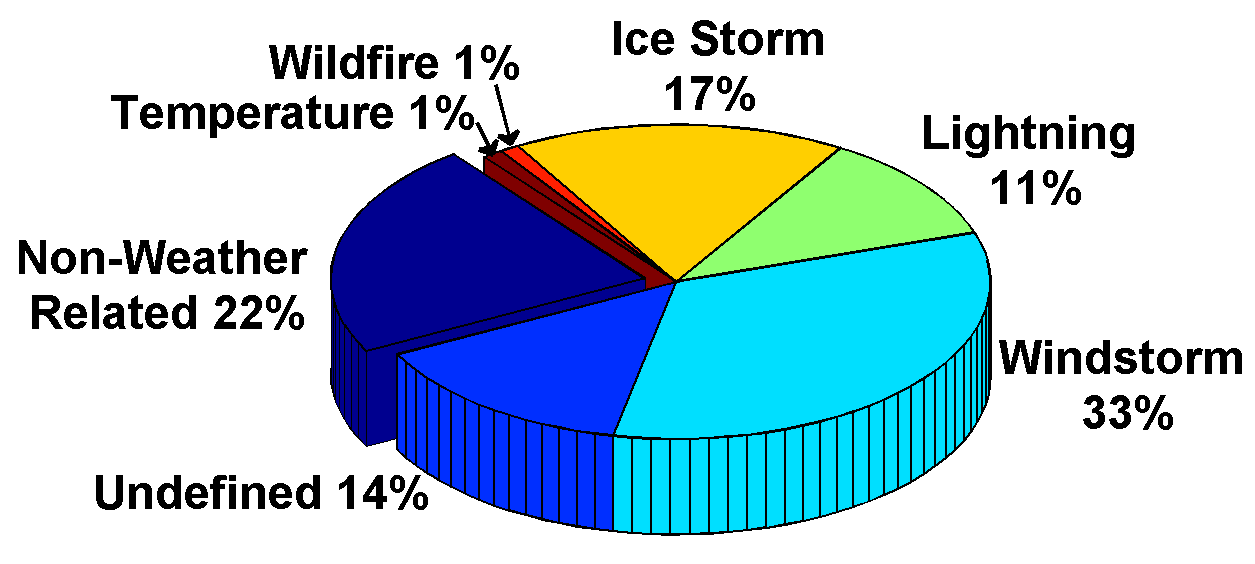}
\caption{The distribution of causes for power interruptions in distribution networks: 178 million customers (meters) collected from 1992 to 2011 in the United States}
\label{fig:perc}
\end{figure}

The power interruptions in distribution networks can be caused by a wide range of factors including equipment failures, animals, trees, and human errors \cite{BOOK1,TTT1}. However, weather conditions, such as windstorm, lightning, ice storm, and high temperature, are usually the most important causes of the distribution power interruptions \cite{Re3,Wea1,Wea2}. According to the reliability data collected by the United States Department of Energy (DOE) \cite{Re31} from 1992 to 2011, the weather conditions contribute more than 64\% of the total number of distribution power interruptions, as depicted in Fig.~\ref{fig:perc}. As a result, the accurate reliability forecasting based on time series of weather condition data is both challenging and desirable for the smart grid distribution system design and operation.

In the last decades, a wealth of researchers have investigated the effects of extreme weather conditions, such as floods, hurricanes, and ice storms, on the power grid reliability performance \cite{Con2,J1,J2,J3}. In \cite{Con2}, a three-state weather model was formulated for the predictive reliability assessment of the electric distribution systems, where the system failure rate was analyzed based on extreme weather conditions. The potential effects of extreme climate changes on power system component reliability were reviewed in \cite{J1}, and the mitigation framework was outlined for boosting the resilience of electrical grids. In \cite{J2}, a mathematical framework was presented to assess the risk of extreme weather cases on the power systems, where the performances of system protection devices were evaluated under extreme events. Additionally, a risk-based defensive islanding algorithm was proposed in \cite{J3} for boosting the power grid resilience to extreme weather events. However, all of these works only consider extreme weather conditions. Although severe weather events can cause large amounts of power interruptions, it is not common to consider these events under normal operation conditions and the major of electric customers' interruptions happen under normal weather conditions.

Recently, a series of efforts have been proposed in \cite{Con3,Arif1,Arif2} to analyze the relationship between the number of power interruptions on electric distribution networks and common weather parameters, such as temperature, wind, air pressure, and lightning. In \cite{Con3}, the different common weather clusters were formulated through the self-organization map clustering analysis, and each cluster of weather parameters was associated with average and standard deviation of the different reliability variables such as the number of power interruptions and system reliability metrics. A weather-based smart grid reliability assessment framework was proposed utilizing the Boolean driven Markov process in \cite{Arif1}, which introduced a method to assess the system reliability performance under variable weather conditions. In \cite{Arif2}, the polynomial regression analysis was introduced to analyze the distribution network response based on various weather parameters, and the number of distribution power interruptions was predicted based on the total sum of the regression model of each weather parameter.

Since the power interruptions related with common weather conditions are essentially the result of combined action of many factors, the power interruption prediction only based on statistical models might be compromising due to the various effects of different weather parameters. In this paper, a hybrid framework integrating multilayer perceptrons (MLPs) and parametric regression models is proposed for forecasting the daily numbers of power interruptions in smart grid distribution networks using time series of common weather data. The main contributions of this paper contain: \textbf{(i)} both polynomial and exponential regression models are implemented to analyze nonlinear relationship between power interruptions and common weather parameters; \textbf{(ii)} derived regression models are integrated as inputs to formulate a MLP neural network to predict the number of power interruptions instead of directly summing together; \textbf{(iii)} a modified extreme learning machine (ELM) based hierarchical algorithm is proposed for training the formulated MLP model using real-time monitored power interruptions data from an electric utility in Florida and common weather data from National Climatic Data Center (NCDC); and \textbf{(iv)} sensitivity analysis is implemented to analyze the various impacts of different common weather parameters on the number of power interruptions.

The rest of this paper is organized as follows. Section 2 introduces the reliability metrics and weather parameters collected for analysis. Section 3 develops the parametric regression models between the reliability metrics and various weather parameters. Section 4 formulates a MLP model for forecasting the daily numbers of power interruptions and introduces a modified ELM based algorithm for training the formulated model. In Section 5, the proposed framework is evaluated, and the sensitivity of each weather parameter is analyzed. Section 6 concludes the paper and outlines the future work.

\section{Reliability metrics and weather data collection and preprocessing}

\begin{figure}[tbp]
\centering
  \includegraphics[width=8cm]{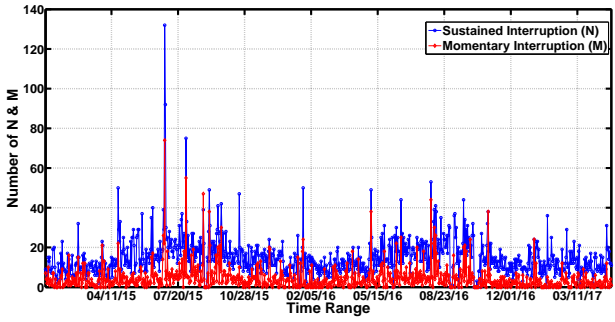}
\caption{Daily numbers of sustained interruptions ($N$) and momentary interruptions ($M$) collected from Jan. 1st 2015 to Apr. 30th 2017 in one utility management area}
\label{fig:NM}       
\end{figure}

In order to analyze the impacts of common weather conditions on the reliability performances of the smart grid distribution networks, a series of reliability metrics are collected from an electric utility in the United States, serving approximately 10 million people across nearly half of the state of Florida. The reliability metrics collected from the electric utility are comprised of the daily numbers of \emph{sustained interruption} ($N$), \emph{momentary interruption} ($M$), \emph{System Average Interruption Duration Index} (SAIDI), \emph{System Average Interruption Frequency Index} (SAIFI), \emph{Momentary Average Interruption Frequency Index} (MAIFI), and \emph{Customer Momentary Experience} (CME). The numbers of $N$ and $M$ play a key point in reliability analysis, and other reliability metrics can be calculated based on these values \cite{Metric1}. Therefore, $N$ and $M$ in one utility management area are selected for reliability analysis in this paper, and a sample of the daily numbers of $N$ and $M$ is shown in Fig. \ref{fig:NM}.

Weather data is mostly collected from NCDC, which provides monthly, daily, and even hourly normal weather data summaries. In addition, a source of weather data like lightning is provided by the control center of the electric utility, which installs its own weather stations centrally located in various management areas. The important weather characteristics contain \emph{temperature}, \emph{precipitation}, \emph{snow}, \emph{air pressure}, \emph{wind speed}, and \emph{lightning}, and all of these data are collected for each utility management area in order to analyze their impacts on the daily numbers of $N$ and $M$.

The power interruption data can be classified into interruptions with exclusion data and without exclusion data for an entire day. In order to analyze the combined effect of common weather conditions on the system reliability performance, the daily numbers of $N$ and $M$ are selected without exclusion data to avoid extreme weather conditions and transmission outages. On the other hand, the weather data is collected hourly to model the data more precisely and improve the location of the weather source. In the following section, the preprocessed power interruptions and weather data will be used for formulating the parametric models to evaluate the effect of each weather parameter on the daily numbers of $N$ and $M$.

\section{Parametric regression models between reliability metrics and common weather data}

In this section, \emph{parametric regression analysis} is introduced to analyze the nonlinear relationship between the number of sustained interruptions and each common weather parameter in one distribution management area. Similarly, the regression analysis can be expanded for the number of momentary interruptions.

For each utility management area, assume that $\boldsymbol{N}$ is the vector of the daily sustained interruptions in one time period, and $\boldsymbol{x}$ is a common weather parameter vector in corresponding time series, the relationship between the two can be represented as: $\boldsymbol{N}=f(\boldsymbol{x},\boldsymbol{\beta})+\boldsymbol{\varepsilon}$, where $f(\cdot)$ is the relationship function; $\boldsymbol{\beta}$ is the vector of estimation parameters; and $\boldsymbol{\varepsilon}$ are the unobserved random error satisfying $\varepsilon\sim{N(0,\delta^2)}$, where $\delta$ is the standard deviation. Two most popular regression models can be implemented for this analysis including polynomial and exponential regression. For polynomial regression with the $n$-th degree, the relationship function can be defined as:
\begin{equation}\label{eq:T2}
f(\boldsymbol{x},\boldsymbol{\beta}^\textrm{pol})=\beta^\textrm{pol}_0+\beta^\textrm{pol}_1{\boldsymbol{x}}+\beta^\textrm{pol}_2{\boldsymbol{x}^2}+\cdot\cdot\cdot+\beta^\textrm{pol}_n{\boldsymbol{x}^n}
\end{equation}
Correspondingly, for two-term exponential regression, the relationship function can be determined by:
\begin{equation}\label{eq:T3}
f(\boldsymbol{x},\boldsymbol{\beta}^\textrm{ex})=\beta^\textrm{ex}_0+\beta^\textrm{ex}_1\exp(\beta^\textrm{ex}_2{\boldsymbol{x}})+\beta^\textrm{ex}_3\exp(\beta^\textrm{ex}_4{\boldsymbol{x}})
\end{equation}

Commonly, based on the sustained power interruptions and weather parameters collected for $N_\textrm{total}$ days: $(\boldsymbol{N}_k,\boldsymbol{x}_k)$, $k = 1,...,N_\textrm{total}$, the least square method~\cite{J4} can be adopted to estimate the optimal estimation parameter $\hat{\boldsymbol{\beta}}$, and the estimation can be expressed by:
\begin{equation}\label{eq:T4}
\sum_{k=1}^{N_\textrm{total}}(\boldsymbol{N}_k-f(\boldsymbol{x}_k,\hat{\boldsymbol{\beta}}))=\min_{\boldsymbol{\beta}}\sum_{k=1}^{N_\textrm{total}}(\boldsymbol{N}_k-f(\boldsymbol{x}_k,{\boldsymbol{\beta}}))
\end{equation}
The goodness-of-fit of polynomial regression with the $n$-th degree, $n=1,2,3$, and two-term exponential regression for each common weather parameter including \emph{temperature}, \emph{wind speed}, \emph{precipitation}, \emph{air pressure}, and \emph{lightning} is listed in Table \ref{SSE}, where each regression model performance is evaluated through the error metrics including \emph{Sum of Squares Due to Error} (SSE), \emph{R-square}, \emph{Degrees of Freedom Adjusted R-Square}, and \emph{Root Mean Squared Error} (RMSE) defined:

\begin{table*}[]
\centering
\caption{The goodness-of-fit of polynomial and exponential regression models for each weather parameter}
\label{SSE}
\resizebox{\textwidth}{!}{
\begin{tabular}{@{}llllll@{}}
\toprule
\textbf{Weather Parameter} & \textbf{Performance} & \textbf{Polynomial ($1$)} & \textbf{Polynomial ($2$)} & \textbf{Polynomial ($3$)} & \textbf{Exponential ($2$)} \\ \shline
\multicolumn{1}{c}{\multirow{4}{*}{\textbf{Maximum}}} & \multicolumn{1}{l|}{SSE} & 7.13E+04   & 6.68E+04  & \textbf{6.63E+04} & 7.42E+04 \\
\multicolumn{1}{c}{\multirow{4}{*}{\textbf{Temperature}}} & \multicolumn{1}{l|}{R-square} & 0.09367    & 0.1517    & \textbf{0.1574}   & 0.05683  \\
\multicolumn{1}{c}{} & \multicolumn{1}{l|}{Adjusted R-square} & 0.0926     & 0.1497    & \textbf{0.1544}   & 0.05349  \\
\multicolumn{1}{c}{} & \multicolumn{1}{l|}{RMSE} & 9.171      & 8.877     & \textbf{8.853}    & 9.366    \\ \midrule
\multicolumn{1}{c}{\multirow{4}{*}{\textbf{Minimum}}} & \multicolumn{1}{l|}{SSE} & 7.47E+04   & 7.39E+04  & \textbf{7.33E+04} & 7.38E+04 \\
\multicolumn{1}{c}{\multirow{4}{*}{\textbf{Temperature}}} & \multicolumn{1}{l|}{R-square}          & 0.05044    & 0.06093   & \textbf{0.06798}  & 0.06238  \\
\multicolumn{1}{l}{} & \multicolumn{1}{l|}{Adjusted R-square} & 0.04932    & 0.05872   & \textbf{0.06467}  & 0.05906  \\
\multicolumn{1}{l}{} & \multicolumn{1}{l|}{RMSE}              & 9.387      & 9.34      & \textbf{9.311}    & 9.338    \\ \midrule
\multicolumn{1}{c}{\multirow{4}{*}{\textbf{Average}}} & \multicolumn{1}{l|}{SSE}               & 7.30E+04   & 7.06E+04  & \textbf{7.05E+04} & 7.06E+04 \\
\multicolumn{1}{c}{\multirow{4}{*}{\textbf{Temperature}}} & \multicolumn{1}{l|}{R-square}          & 0.07222    & 0.1031    & \textbf{0.104}    & 0.1028   \\
\multicolumn{1}{l}{} & \multicolumn{1}{l|}{Adjusted R-square} & 0.07113    & 0.101     & \textbf{0.1008}   & 0.09958  \\
\multicolumn{1}{l}{} & \multicolumn{1}{l|}{RMSE}              & 9.278      & 9.128     & \textbf{9.129}    & 9.135    \\ \midrule
\multicolumn{1}{c}{\multirow{4}{*}{\textbf{Heat Degree Days}}} & \multicolumn{1}{l|}{SSE}               & 7.87E+04   & 7.85E+04  & 7.85E+04 & \textbf{7.83E+04} \\
\multicolumn{1}{l}{} & \multicolumn{1}{l|}{R-square}          & 0.0002128  & 0.002496  & 0.002849 & \textbf{0.005259} \\
\multicolumn{1}{l}{} & \multicolumn{1}{l|}{Adjusted R-square} & -0.0009662 & 0.0001408 & -0.0006875  & \textbf{0.001732} \\
\multicolumn{1}{l}{} & \multicolumn{1}{l|}{RMSE}  & 9.632      & 9.626     & 9.63     & \textbf{9.619}    \\ \midrule
\multicolumn{1}{c}{\multirow{4}{*}{\textbf{Cool Degree Days}}} & \multicolumn{1}{l|}{SSE}               & 7.22E+04   & \textbf{7.08E+04}  & \textbf{7.08E+04} & \textbf{7.08E+04} \\
\multicolumn{1}{l}{} & \multicolumn{1}{l|}{R-square} & 0.08255    & \textbf{0.1006}    & \textbf{0.1006}   & \textbf{0.1006}   \\
\multicolumn{1}{l}{} & \multicolumn{1}{l|}{Adjusted R-square} & 0.08147    & \textbf{0.09848}   & \textbf{0.09848}  & \textbf{0.09848}  \\
\multicolumn{1}{l}{} & \multicolumn{1}{l|}{RMSE}              & 9.227      & \textbf{9.141}     & \textbf{9.141}    & \textbf{9.141}    \\ \midrule
\multicolumn{1}{c}{\multirow{4}{*}{\textbf{Precipitation}}} & \multicolumn{1}{l|}{SSE}               & 6.69E+04   & 6.61E+04  & 6.49E+04 & \textbf{6.43E+04} \\
\multicolumn{1}{l}{} & \multicolumn{1}{l|}{R-square} & 0.08588    & 0.0963    & 0.1133   & \textbf{0.1212}   \\
\multicolumn{1}{l}{} & \multicolumn{1}{l|}{Adjusted R-square} & 0.08462    & 0.09381   & 0.1096   & \textbf{0.1175}   \\
\multicolumn{1}{l}{} & \multicolumn{1}{l|}{RMSE}              & 9.596      & 9.547     & 9.464    & \textbf{9.422}    \\ \midrule
\multicolumn{1}{c}{\multirow{4}{*}{\textbf{Air Pressure}}} & \multicolumn{1}{l|}{SSE}               & 7.71E+04   & 7.66E+04  & \textbf{7.57E+04} & 7.60E+04 \\
\multicolumn{1}{l}{} & \multicolumn{1}{l|}{R-square}          & 0.02001    & 0.02647   & \textbf{0.03759}  & 0.03367  \\
\multicolumn{1}{l}{} & \multicolumn{1}{l|}{Adjusted R-square} & 0.01886    & 0.02417   & \textbf{0.03417}  & 0.03024  \\
\multicolumn{1}{l}{} & \multicolumn{1}{l|}{RMSE}              & 9.536      & 9.51      & \textbf{9.461}    & 9.48     \\ \midrule
\multicolumn{1}{c}{\multirow{4}{*}{\textbf{Average}}}  & \multicolumn{1}{l|}{SSE}               & 7.62E+04   & 7.51E+04  & 7.49E+04 & \textbf{7.48E+04} \\
\multicolumn{1}{c}{\multirow{4}{*}{\textbf{Wind Speed}}} & \multicolumn{1}{l|}{R-square}          & 0.03019    & 0.04404   & 0.04623  & \textbf{0.04843}  \\
\multicolumn{1}{l}{} & \multicolumn{1}{l|}{Adjusted R-square} & 0.02904    & 0.04177   & 0.04282  & \textbf{0.04504}  \\
\multicolumn{1}{l}{} & \multicolumn{1}{l|}{RMSE}              & 9.506      & 9.444     & 9.438    & \textbf{9.427}    \\ \midrule
\multicolumn{1}{c}{\multirow{4}{*}{\textbf{Peak}}} & \multicolumn{1}{l|}{SSE} & 7.71E+04   & 7.66E+04  & 7.61E+04 & \textbf{7.44E+04} \\
\multicolumn{1}{c}{\multirow{4}{*}{\textbf{Wind Speed}}} & \multicolumn{1}{l|}{R-square}          & 0.01996    & 0.02649   & 0.03265  & \textbf{0.05445}  \\
\multicolumn{1}{l}{} & \multicolumn{1}{l|}{Adjusted R-square} & 0.0188     & 0.02418   & 0.02921  & \textbf{0.05108}  \\
\multicolumn{1}{l}{} & \multicolumn{1}{l|}{RMSE}              & 9.551      & 9.525     & 9.501    & \textbf{9.393}    \\ \midrule
\multicolumn{1}{c}{\multirow{4}{*}{\textbf{Sustainable}}}  & \multicolumn{1}{l|}{SSE}               & 7.77E+04   & 7.76E+04  & \textbf{7.72E+04} & 7.74E+04 \\
\multicolumn{1}{c}{\multirow{4}{*}{\textbf{Wind Speed}}} & \multicolumn{1}{l|}{R-square}          & 0.01321    & 0.01411   & \textbf{0.01847}  & 0.01641  \\
\multicolumn{1}{l}{} & \multicolumn{1}{l|}{Adjusted R-square} & 0.01205    & 0.01178   & \textbf{0.01499}  & 0.01293  \\
\multicolumn{1}{l}{} & \multicolumn{1}{l|}{RMSE}              & 9.569      & 9.57      & \textbf{9.555}    & 9.565    \\ \midrule
\multicolumn{1}{c}{\multirow{4}{*}{\textbf{Lightning}}} & \multicolumn{1}{l|}{SSE}               & 6.60E+04   & 6.31E+04  & \textbf{6.26E+04} & 6.35E+04 \\
\multicolumn{1}{l}{} & \multicolumn{1}{l|}{R-square}          & 0.1609     & 0.1987    & \textbf{0.2041}   & 0.1937   \\
\multicolumn{1}{l}{} & \multicolumn{1}{l|}{Adjusted R-square} & 0.16       & 0.1968    & \textbf{0.2013}   & 0.1908   \\
\multicolumn{1}{l}{} & \multicolumn{1}{l|}{RMSE}              & 8.824      & 8.628     & \textbf{8.604}    & 8.66     \\ \shline
\multicolumn{1}{c}{\multirow{2}{*}{\textbf{\textit{Remarks:}}}} & \multicolumn{5}{l}{$\textbf{SSE}$ measures the total deviation of the predicted values from the fit to the observed values.} \\
\multicolumn{1}{c}{} & \multicolumn{5}{l}{\textbf{R-square} is the square of the correlation between the response values and predicted response.}\\
\multicolumn{1}{c}{} & \multicolumn{5}{l}{\textbf{Adjusted R-square} is a modified R-squared that has been adjusted for the number of predictors.} \\
\multicolumn{1}{c}{} & \multicolumn{5}{l}{\textbf{RMSE} is the estimate of the standard deviation of the random component in the data.}\\
\shline
\end{tabular}}
\end{table*}

\begin{itemize}
\item $\textbf{SSE}=\sum_{k=1}^{\boldsymbol{N}_\textrm{total}}{\boldsymbol{w}_k(\boldsymbol{x}_k-\boldsymbol{f}_k)^2}$ measures the total deviation of the original values from the regression fit, where $\boldsymbol{f}_k$ is the predicted value derived by the regression model, and $\boldsymbol{w}_k$ is the weighting applied to each data point. SSE closer to $0$ indicates that the regression model has a smaller fitting error, which is more useful for prediction.
\item $\textbf{R-square}=1-\frac{[\sum_{k=1}^{\boldsymbol{N}_\textrm{total}}{\boldsymbol{w}_k(\boldsymbol{x}_k-f_k)^2}]}{[\sum_{k=1}^{\boldsymbol{N}_\textrm{total}}{w_k(\boldsymbol{x}_k-\boldsymbol{x}_{\textrm{av}})^2}]}$ denotes the square of the correlation between the original and predicted values, where $\boldsymbol{x}_{av}$ is the mean of the original data. R-square has the unit interval $[0,1]$, where a value closer to 1 represents that a greater proportion of variance is accounted by the regression model.
\item $\textbf{Adjusted R-square}=1-\frac{[\sum_{k=1}^{\boldsymbol{N}_\textrm{total}}{\boldsymbol{w}_k(\boldsymbol{x}_k-\boldsymbol{f}_k)^2}](\boldsymbol{N}_\textrm{total}-1)}{[\sum_{k=1}^{\boldsymbol{N}_\textrm{total}}{\boldsymbol{w}_k(\boldsymbol{x}_k-\boldsymbol{x}_{\textrm{av}})^2}](v)}$ is a modified type of R-square, where the residual degrees of freedom $v$ is determined by $\boldsymbol{N}_\textrm{total}$ minus the number of fitted coefficients. The adjusted R-square statistic can take on any value less than or equal to 1, where a value closer to 1 implies a better regression fitting.
\item $\textbf{RMSE}=\sqrt[]{SSE/v}$ is correlated with the fit standard error and the standard error of the regression model, which is an estimate of the standard deviation of the random component in the data. Similar with SSE, a RMSE value closer to $0$ indicates that the regression model is more useful for prediction.
\end{itemize}
In the subsequent subsections, the regression models between $\boldsymbol{N}$ and various weather parameters are detailed described.


\subsection{Temperature}
Temperature is an important weather parameter impacting on the reliability performance of smart grid distribution networks. An increase of the power interruptions can be caused at relatively low or high temperatures, since the electricity demand will increase due to the heating/cooling requirements of customers. Moreover, high air temperatures may strain power infrastructure devices and reduce transmission capacity~\cite{J5}. In order to achieve the relationship between temperature and $\boldsymbol{N}$, a series of temperature characteristics including dry-bulb maximum temperature $T_{\textrm{max}}$, average temperature $T_{\textrm{ave}}$, minimum temperature $T_{\textrm{min}}$, heat degree days $H$, and cool degree days $C$ are selected.

\begin{figure*} [tbp] \centering
\subfigure[Maximum, average, and minimum temperature] {\label{fig:a}\includegraphics[width=8.5cm]{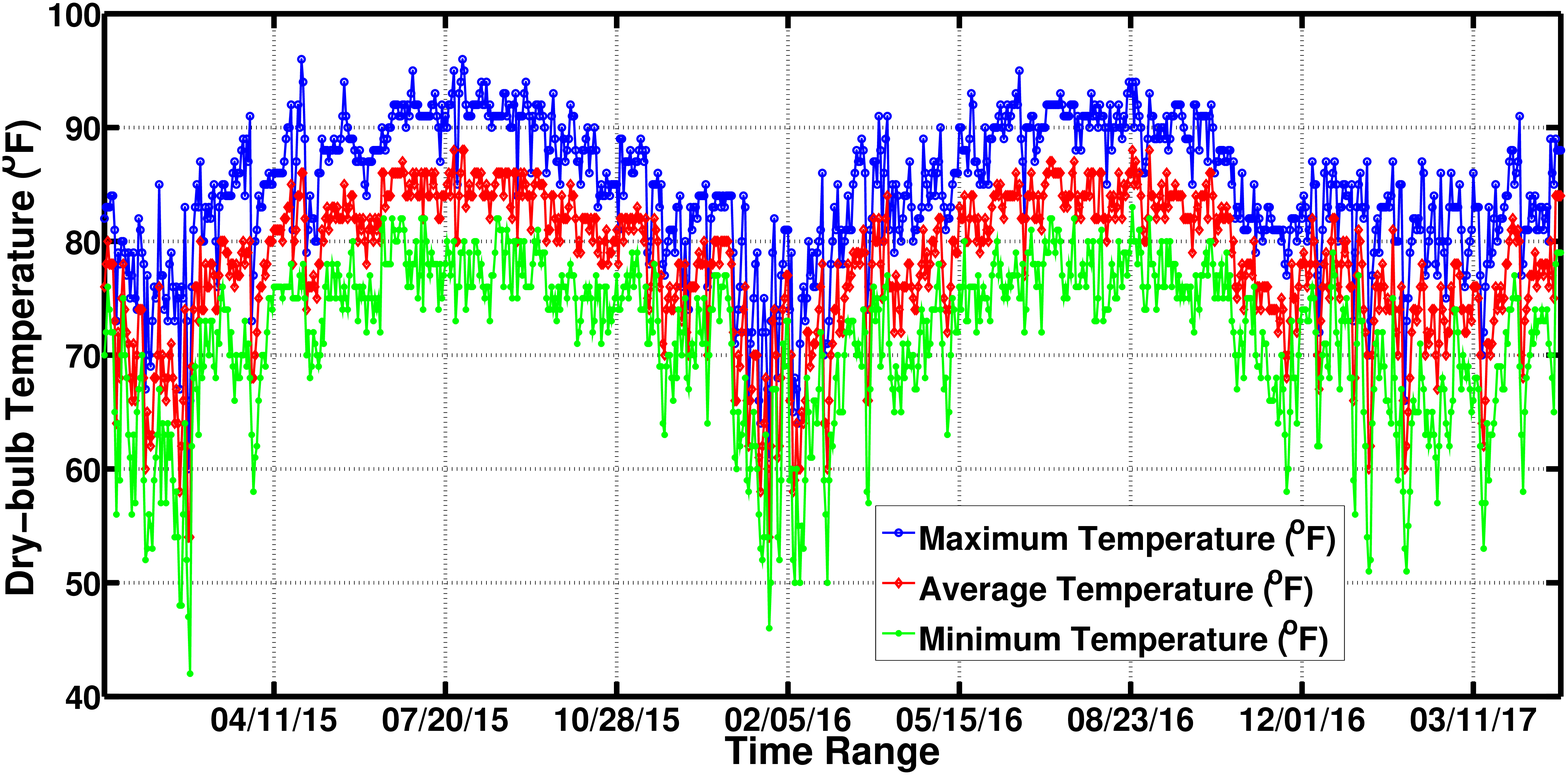}}
\subfigure[Heat and cool degree days] { \label{fig:b}\includegraphics[width=8.5cm]{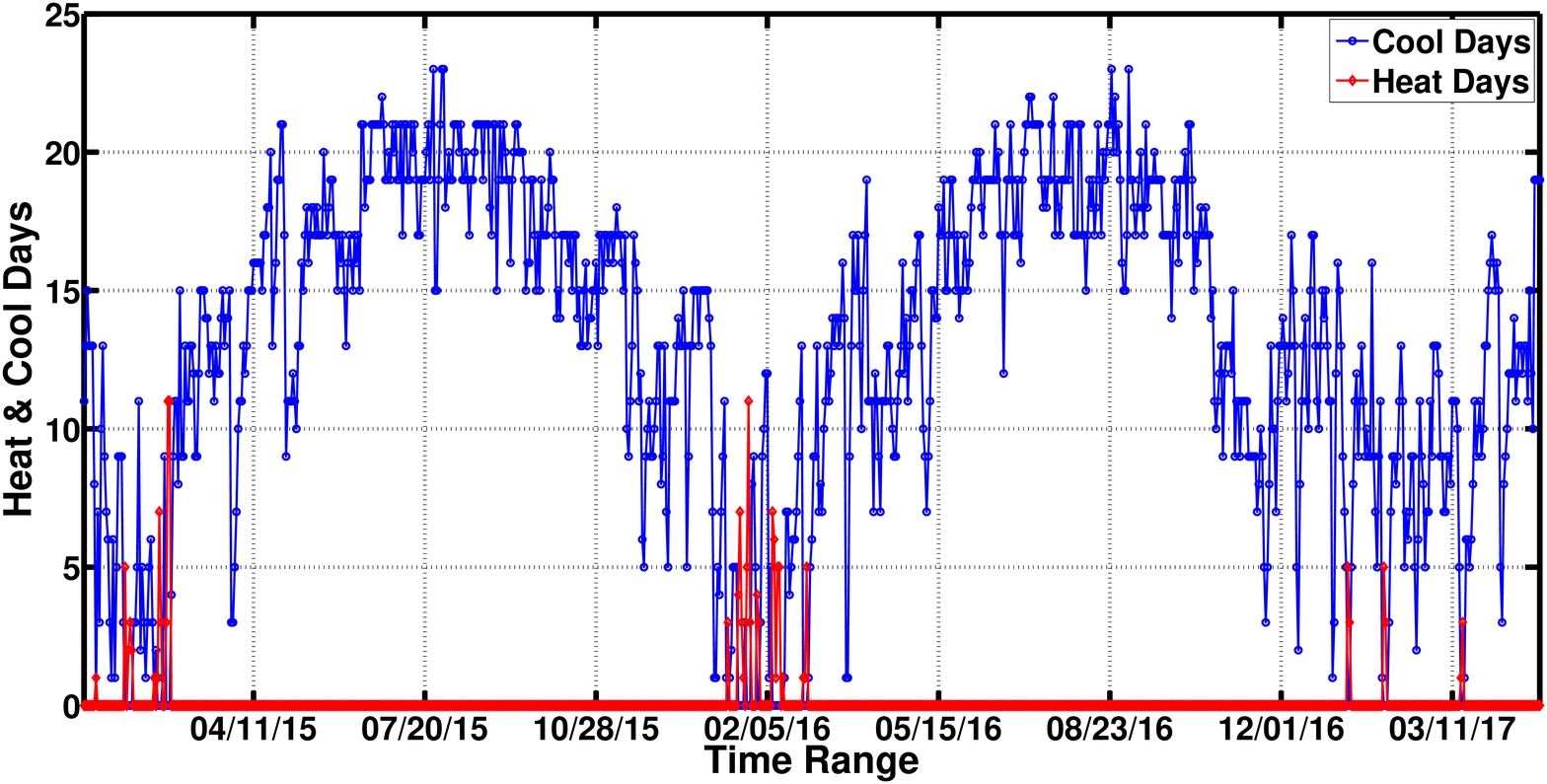}}

\subfigure[Regression analysis for maximum temperature] {\label{fig:c}\includegraphics[width=8.5cm]{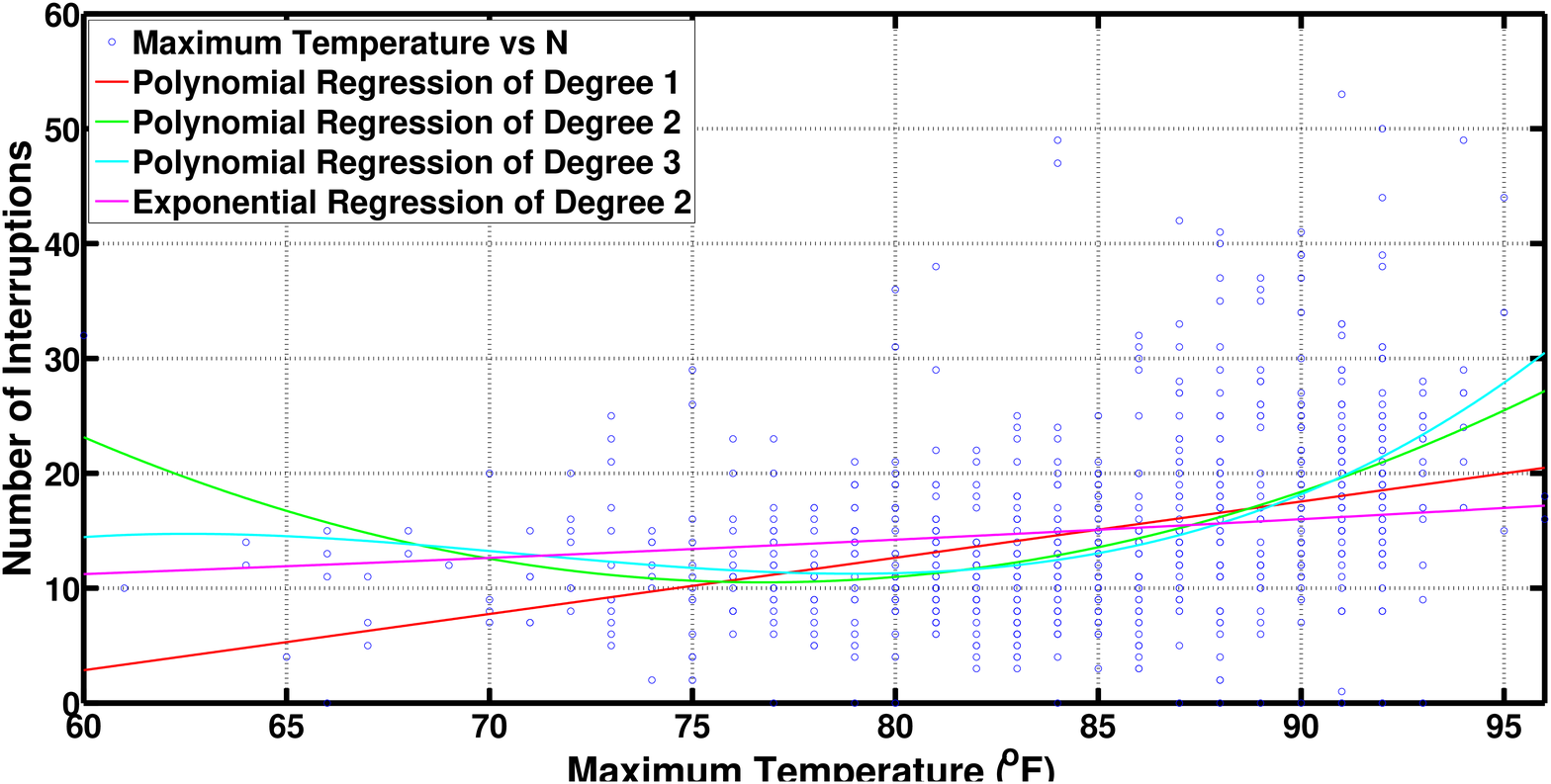}}
\subfigure[Regression analysis for cool degree days] {\label{fig:d}\includegraphics[width=8.5cm]{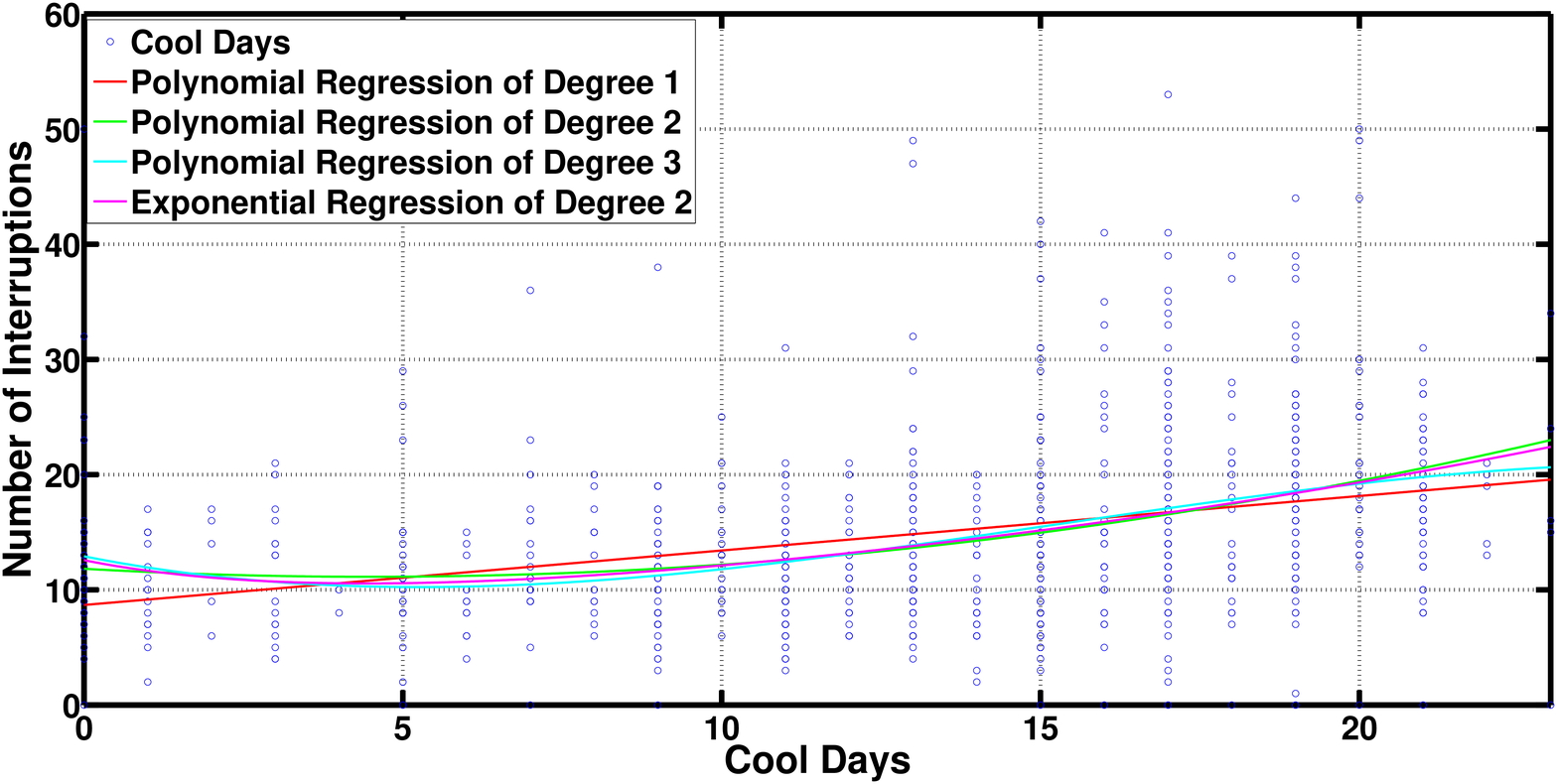}}
\caption{ (a) Maximum temperature $T_{\textrm{max}}$, average temperature $T_{\textrm{ave}}$, and minimum temperature $T_{\textrm{min}}$ (b) Heat degree days $H$ and cool degree days $C$ (c) The polynomial and exponential regression for analyzing the number of $N_{T_{\textrm{max}}}$ respond to maximum temperature $T_{\textrm{max}}$ (d) The polynomial and exponential regression for analyzing the number of $N_{C}$ respond to cool degree days $C$}
\vspace{-0.3cm}
\label{fig:Temp}
\end{figure*}

For one utility management area, Fig. \ref{fig:Temp}(a) shows the variations of $T_{\textrm{max}}$, $T_{\textrm{ave}}$, and $T_{\textrm{min}}$, and Fig. \ref{fig:Temp}(b) displays the variations of $H$ and $C$. Polynomial regression with the $n$-th degree, $n=1,2,3$, and two-term exponential regression are implemented for analyzing the relationship between each temperature characteristic and $\boldsymbol{N}$. Based on the goodness-of-fit results in Table \ref{SSE}, for $T_{\textrm{max}}$, $T_{\textrm{ave}}$, and $T_{\textrm{min}}$, polynomial regression with $3$-th degree has a better performance with less SSE and RMSE, and greater R-square. Exponential regression with $2$ terms has a better performance for $H$, while, for $C$, polynomial regression with $2$-rd and $3$-th degree and exponential regression with $2$ terms have similar performances in modeling fitting. Fig. \ref{fig:Temp}(c) and (d) present the polynomial and exponential regression models adapted for analyzing $T_{\textrm{max}}$ and $C$, respectively. The relationship function between $N_{T_{\textrm{max}}}$ and $T_{\textrm{max}}$ can be expressed as: $N_{T_{\textrm{max}}}=\beta^{T_{\textrm{max}}}_0+\beta^{T_{\textrm{max}}}_1{{T_{\textrm{max}}}}+\beta^{T_{\textrm{max}}}_2{{T_{\textrm{max}}}^2}+\beta^{T_{\textrm{max}}}_3{{T_{\textrm{max}}}^3}$. Correspondingly, the relationship function between $N_{C}$ and $C$ can be determined by: $N_{C}=\beta^{C}_0+\beta^{C}_1\exp(\beta^{C}_2{C})+\beta^{C}_3\exp(\beta^{C}_4{C})$.

\subsection{Wind speed}
Wind also plays a key role in the reliability analysis of smart grid distribution networks. The power of wind is directly proportional to the cube of the wind speed. If the wind reaches high speeds, it can cause damages to distribution networks such as entire trees blowing over into power lines, which results in broken conductors, broken crossarms, broken insulators, broken poles, and leaning poles \cite{W1}. There are three factors which attribute to the severity of wind speeds: peak wind speed $W_{\textrm{pea}}$, average wind speed $W_{\textrm{ave}}$, and sustained wind speed $W_{\textrm{sus}}$.

\begin{figure*} [tbp] \centering
\subfigure[Peak, average, and sustained wind speed] {\label{fig:a}\includegraphics[width=8.5cm]{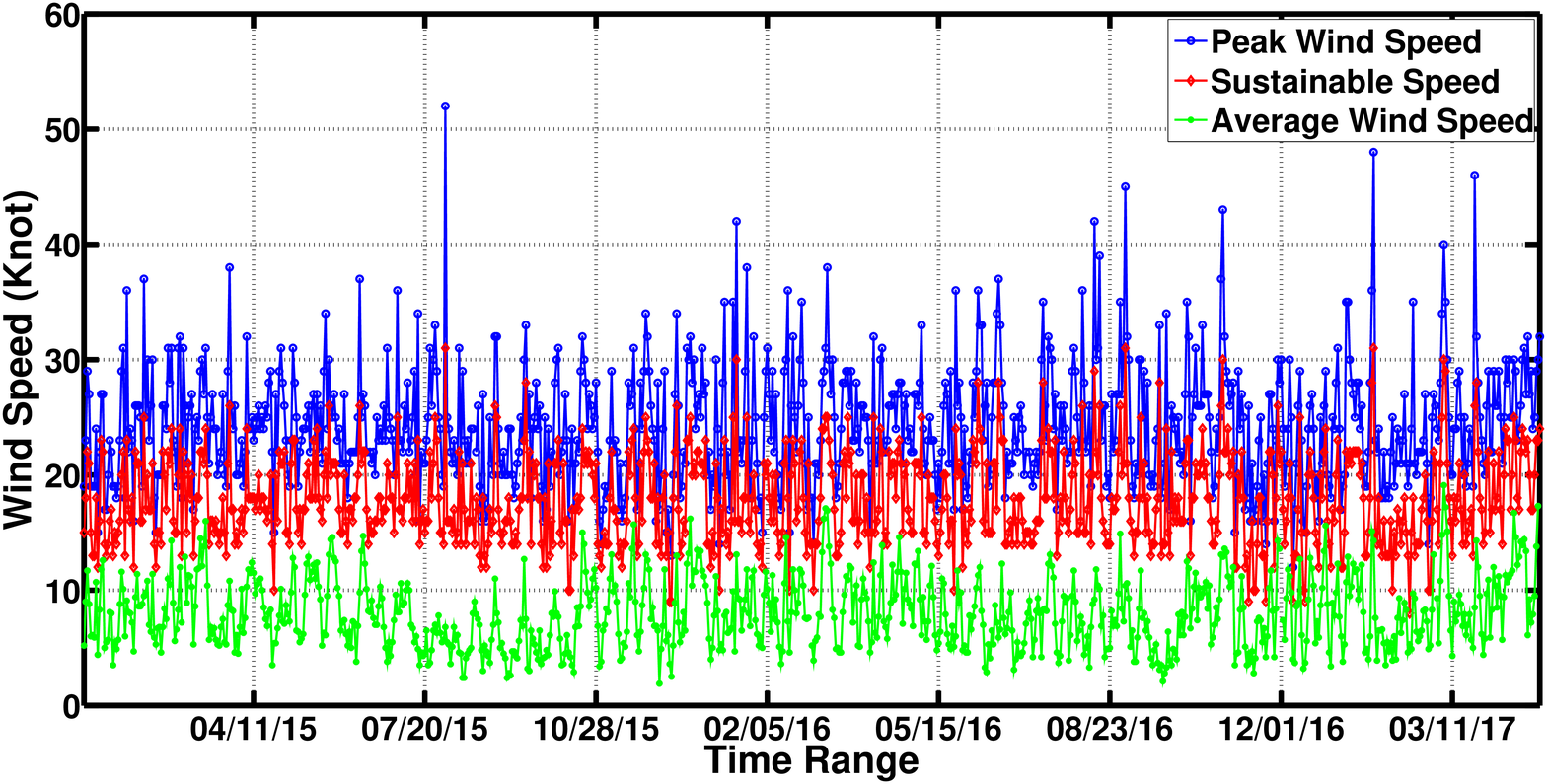}}
\subfigure[Regression analysis for peak wind speed] { \label{fig:b}\includegraphics[width=8.5cm]{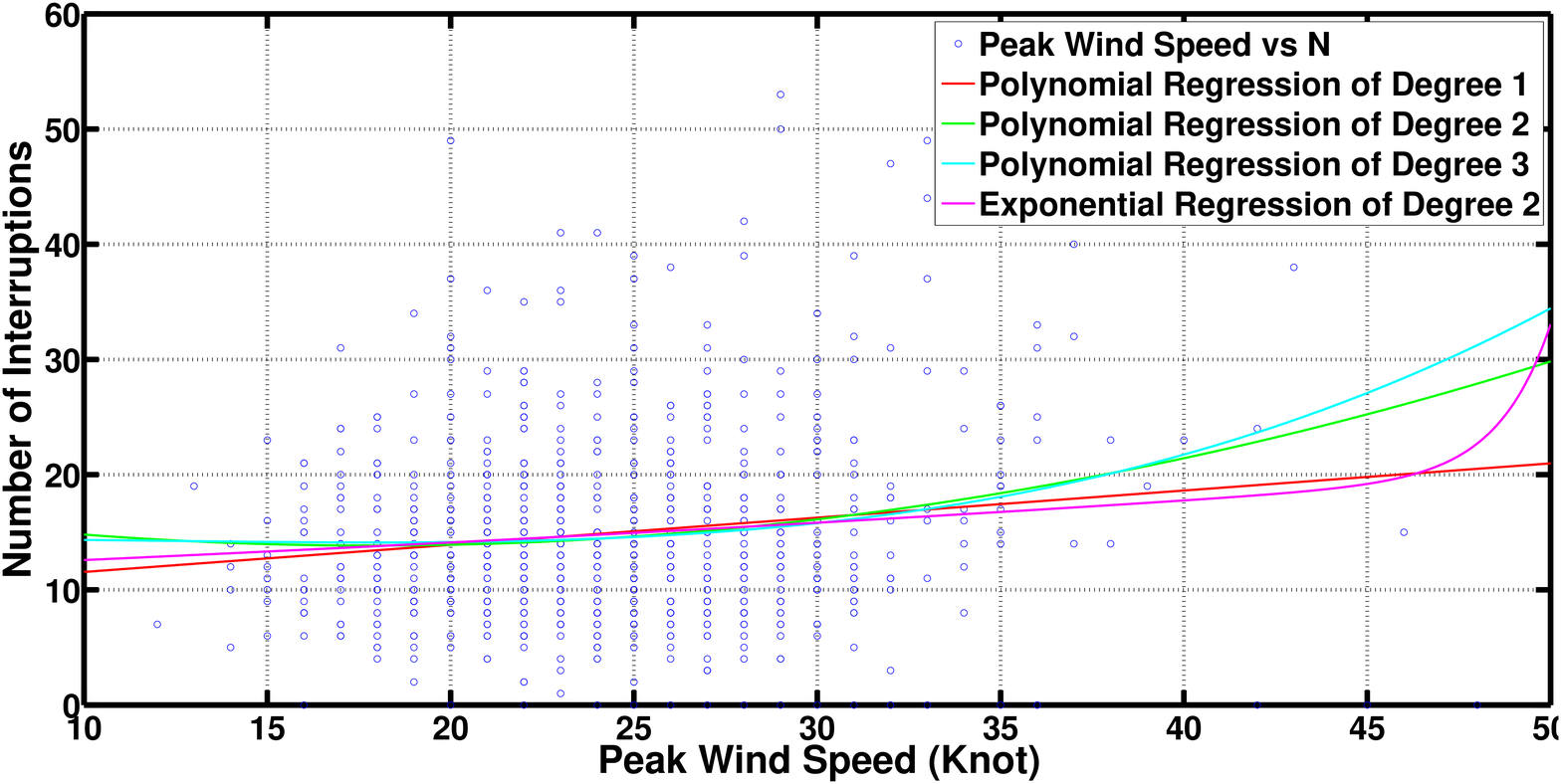}}
\caption{(a) Peak wind speed $W_{\textrm{pea}}$, average wind speed $W_{\textrm{ave}}$, and sustained wind speed $W_{\textrm{sus}}$ (b) The polynomial and exponential regression for analyzing the number of $N_{W_{\textrm{pea}}}$ respond to peak wind speed $W_{\textrm{pea}}$}
\vspace{-0.3cm}
\label{fig:Wind}
\end{figure*}

Fig. \ref{fig:Wind}(a) plots the variation of $W_{\textrm{pea}}$, $W_{\textrm{ave}}$, and $W_{\textrm{sus}}$ for one utility management area, and Fig. \ref{fig:Wind}(b) shows the polynomial and exponential regression models fitted for analyzing the number of $N_{W_{\textrm{pea}}}$ corresponding to $W_{\textrm{pea}}$. Based on the goodness-of-fit results in Table \ref{SSE}, exponential regression with $2$ terms has a better performance for $W_{\textrm{pea}}$ and $W_{\textrm{ave}}$, while $W_{\textrm{sus}}$ has a better performance in polynomial regression with $3$-th degree. The relationship function modeling of the effect of $W_{\textrm{pea}}$ on $N_{W_{\textrm{pea}}}$ can be defined as:
$N_{W_{\textrm{pea}}}=\beta^{W_{\textrm{pea}}}_0+\beta^{W_{\textrm{pea}}}_1\exp(\beta^{W_{\textrm{pea}}}_2{W_{\textrm{p}}})+\beta^{W_{\textrm{pea}}}_3\exp(\beta^{W_{\textrm{pea}}}_4{W_{\textrm{pea}}})$.

\subsection{Precipitation}
\begin{figure*} [tbp] \centering
\subfigure[Raining precipitation] {\label{fig:a}\includegraphics[width=8.5cm]{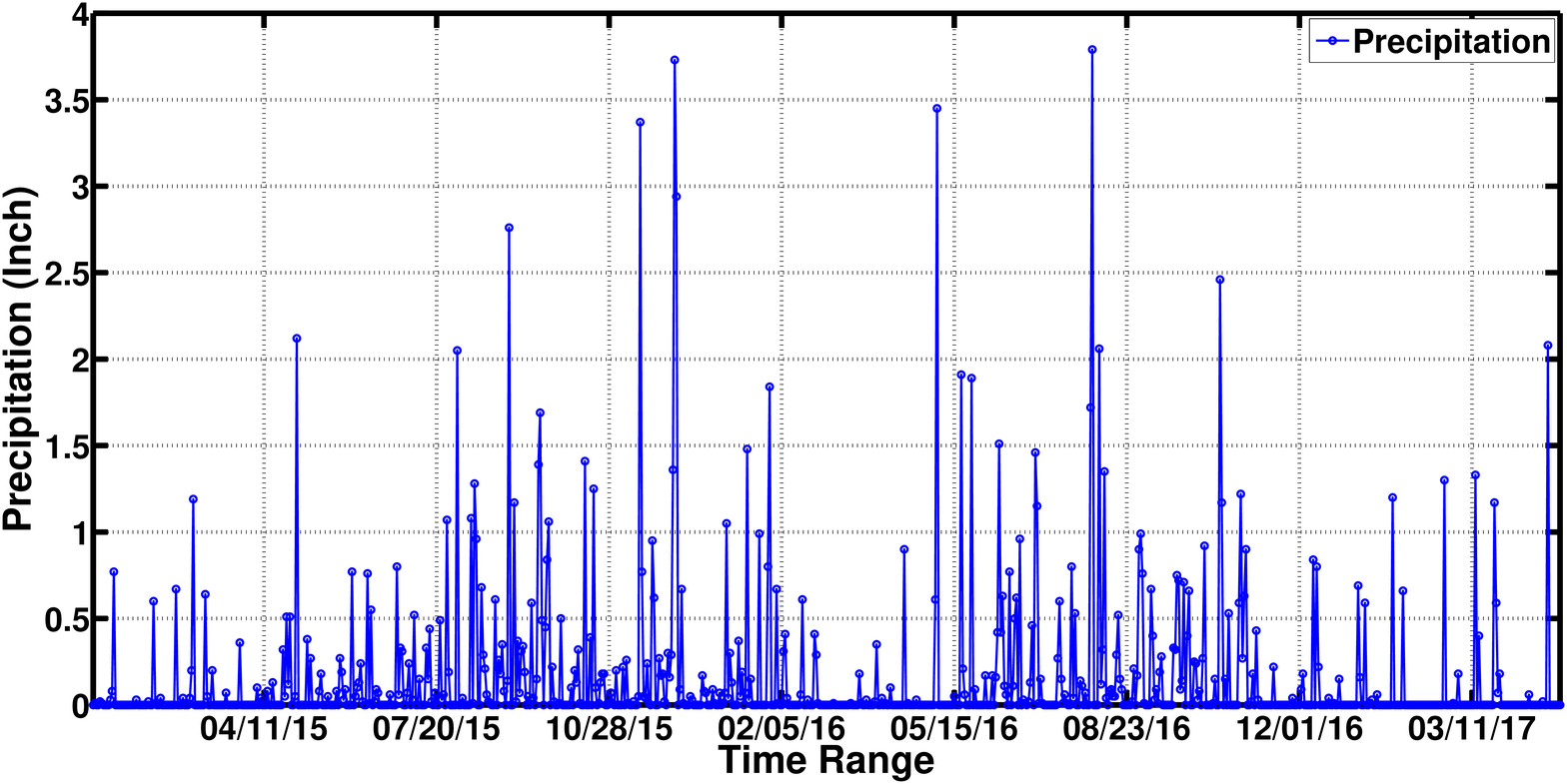}}
\subfigure[Regression analysis for raining precipitation] { \label{fig:b}\includegraphics[width=8.5cm]{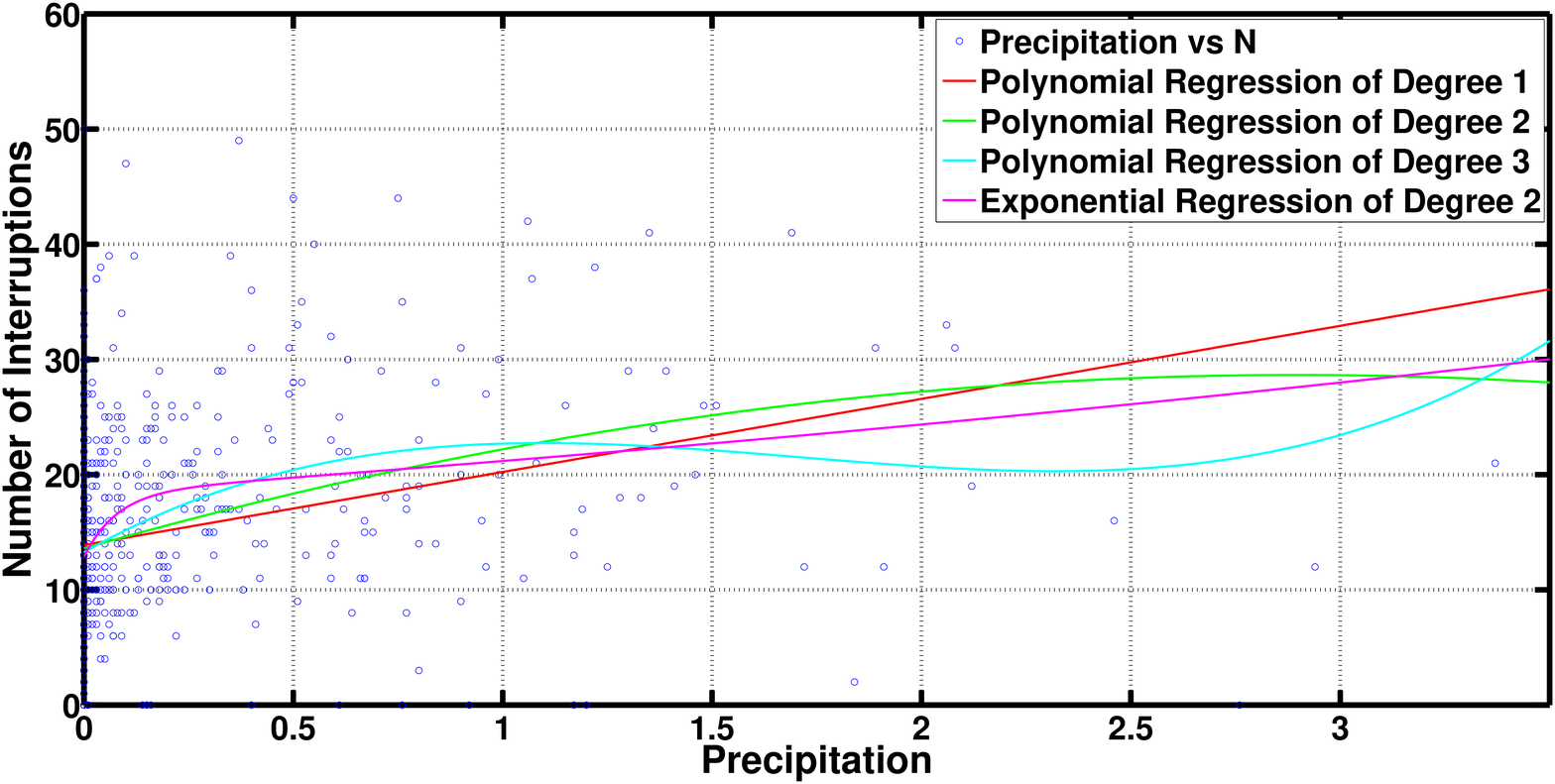}}
\caption{(a) Raining precipitation $P_{\textrm{rain}}$ (b) The polynomial and exponential regression for analyzing the number of $N_{P_{\textrm{rain}}}$ respond to raining precipitation $P_{\textrm{rain}}$}
\vspace{-0.3cm}
\label{fig:Prec}
\end{figure*}

Precipitation is any product of the condensation of atmospheric water vapor that falls under gravity, and two main forms of precipitation contain rain $P_{\textrm{rain}}$ and snow $P_{\textrm{snow}}$. If raining density is large, formerly underground cables, vaults, and manholes may be exposed. Additionally, many power infrastructure equipments may not be sufficient to cater for heavy rain conditions, especially at ultra-high voltage \cite{Pr1}. On the other hand, snow occurs when supercooled rain freezes on contact with tree branches and overhead conductors. Ice buildup on conductors places a heavy physical load on the conductors, which increases the cross-sectional area exposed to the wind \cite{Pr2}. Since little snow falls in Florida, the correlation of raining $P_{\textrm{rain}}$ to $N_{P_{\textrm{rain}}}$ is evaluated in this paper through polynomial and exponential regression models. As shown in Table \ref{SSE}, exponential regression with $2$ terms has a better performance with less SSE and RMSE, and greater R-square for $P_{\textrm{rain}}$. For this purpose, the following equation is developed for modeling the effect of $P_{\textrm{rain}}$ on $N_{P_{\textrm{rain}}}$: $N_{P_{\textrm{rain}}}=\beta^{\textrm{rain}}_0+\beta^{\textrm{rain}}_1\exp(\beta^{\textrm{rain}}_2{P_{\textrm{rain}}})+\beta^{\textrm{rain}}_3\exp(\beta^{\textrm{rain}}_4{P_{\textrm{rain}}})$.
Fig. \ref{fig:Prec}(a) plots the variation of raining precipitation $P_{\textrm{rain}}$ for one utility management area, and Fig. \ref{fig:Prec}(b) shows the polynomial and exponential regression fittings adapted for analyzing the relationship between $P_{\textrm{rain}}$ and $N_{P_{\textrm{rain}}}$.

\begin{figure*} [tbp] \centering
\subfigure[Air pressure] {\label{fig:a}\includegraphics[width=8.5cm]{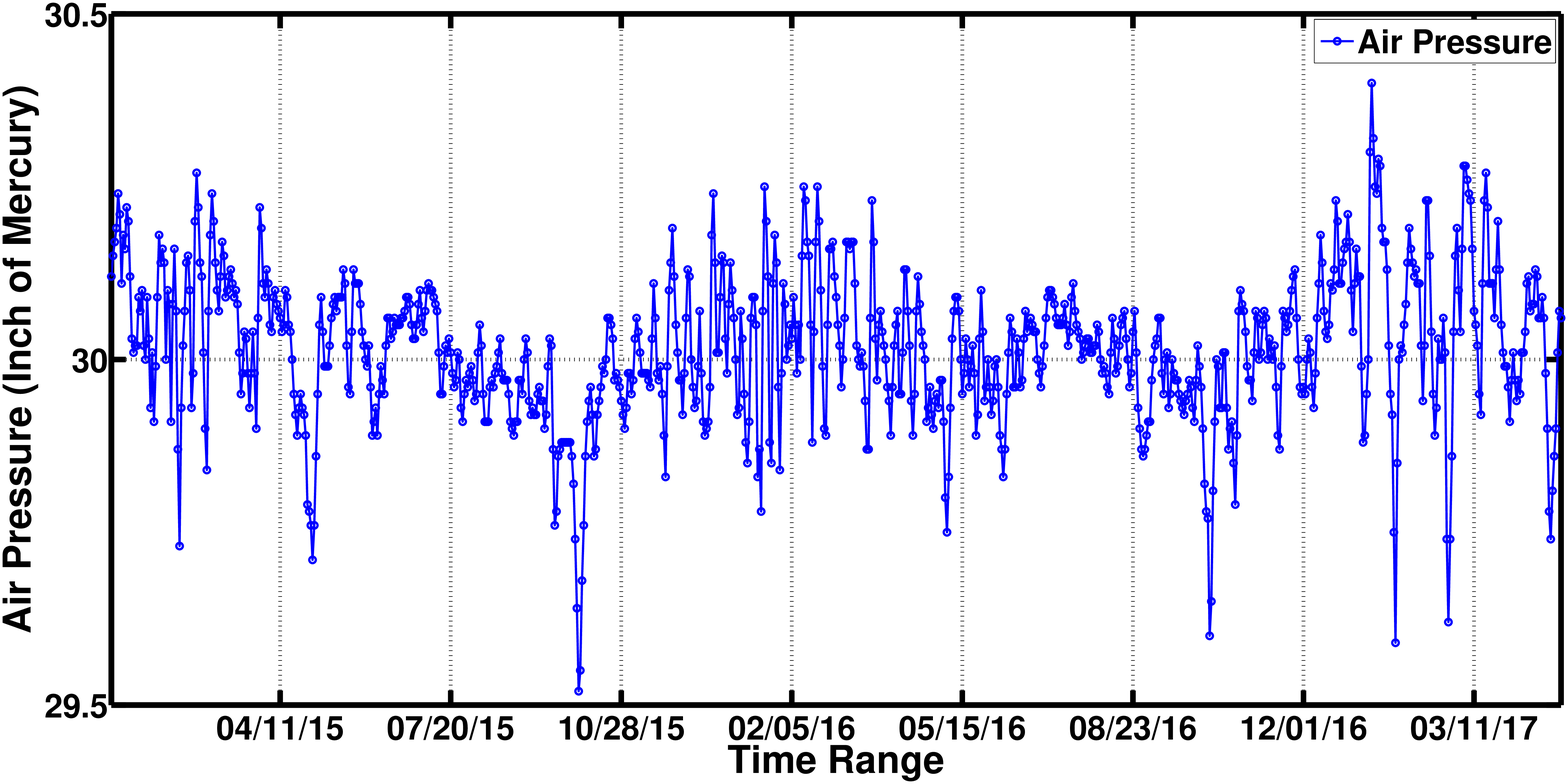}}
\subfigure[Regression analysis for air pressure] { \label{fig:b}\includegraphics[width=8.5cm]{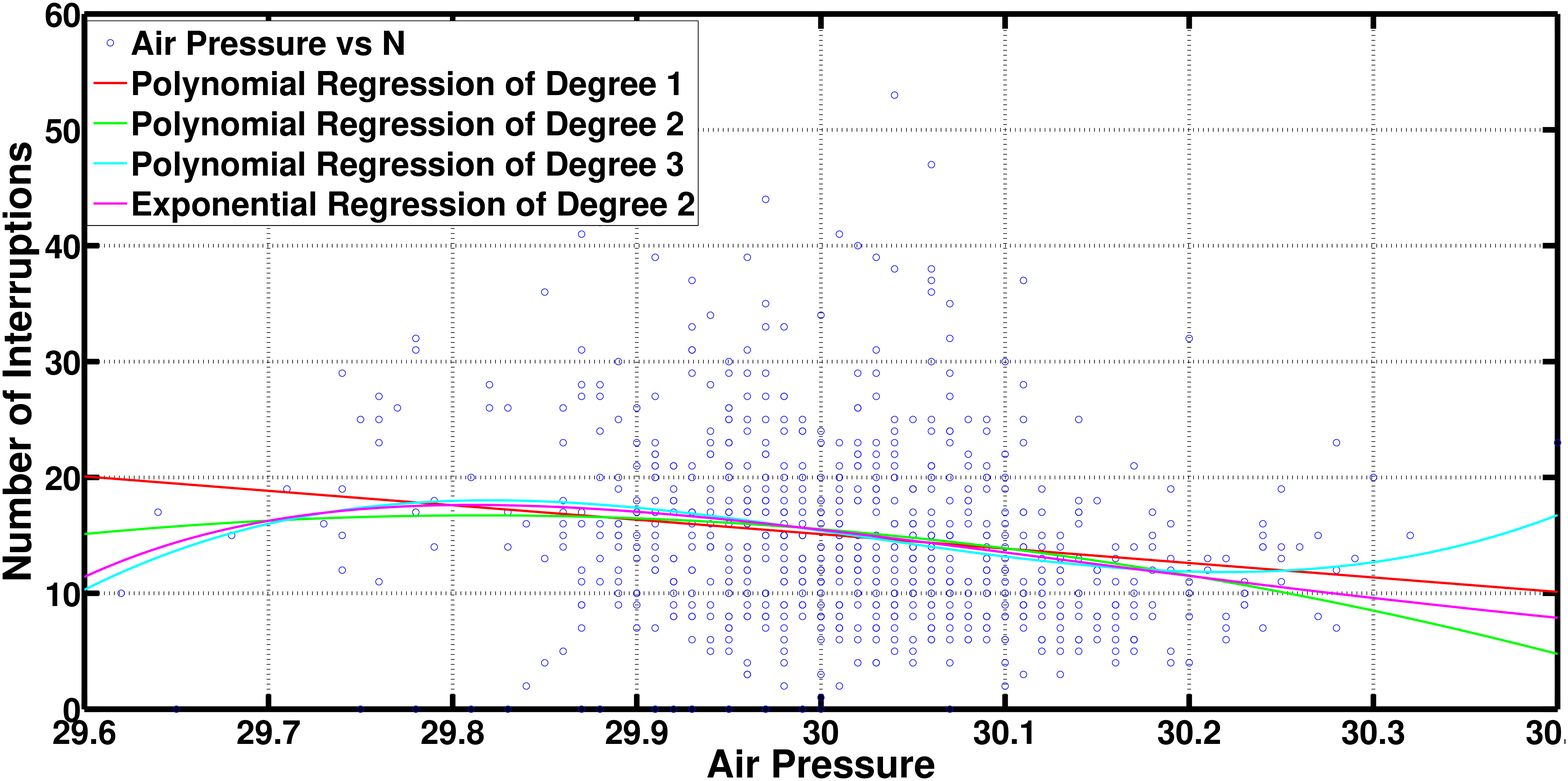}}
\caption{(a) Air pressure $A$ (b) The polynomial and exponential regression for analyzing the number of $N_{A}$ respond to air pressure $A$}
\vspace{-0.3cm}
\label{fig:Air}
\end{figure*}

\begin{figure*} [tbp] \centering
\subfigure[Lightning] {\label{fig:a}\includegraphics[width=8.5cm]{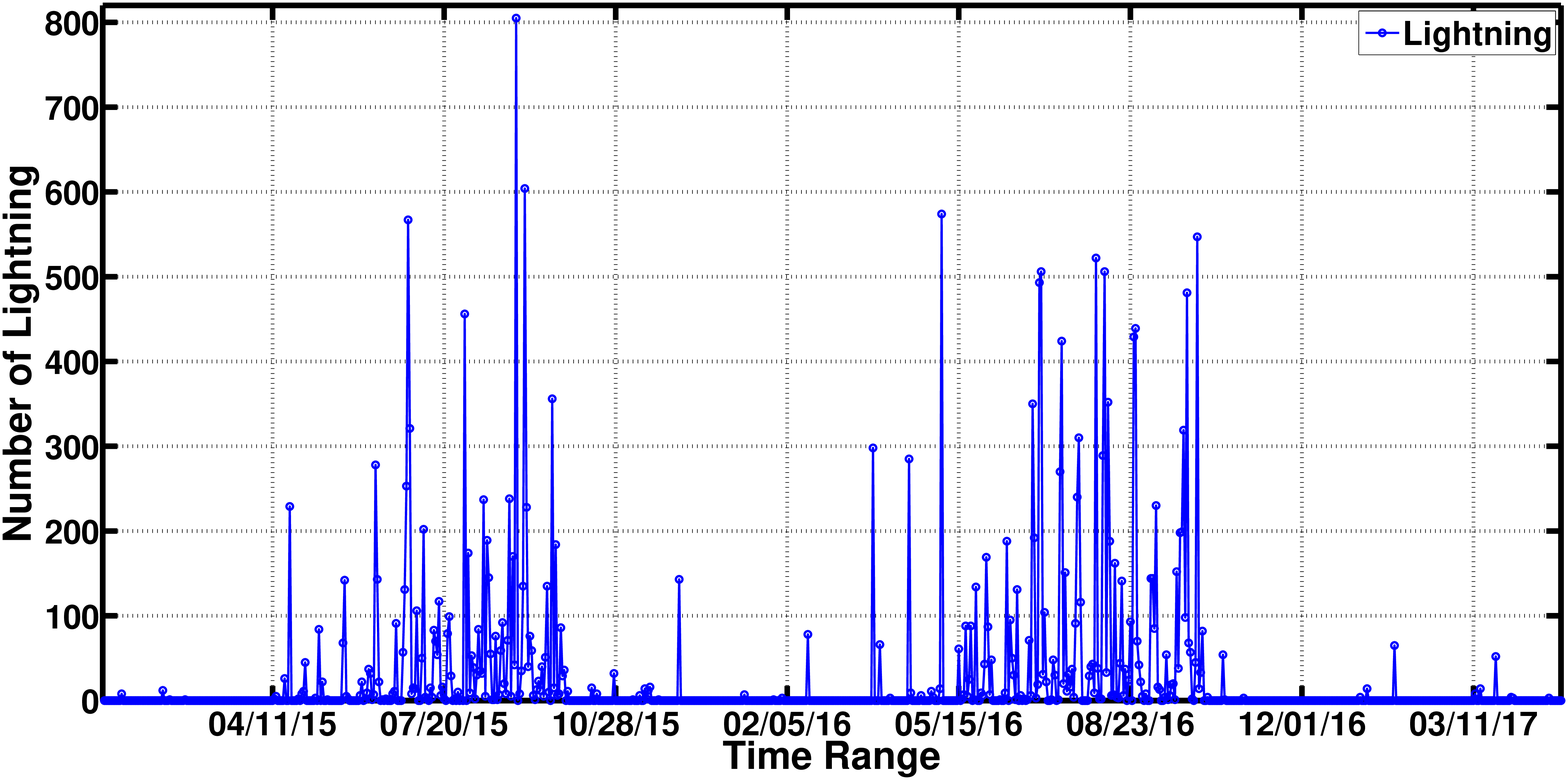}}
\subfigure[Regression analysis for lightning] { \label{fig:b}\includegraphics[width=8.5cm]{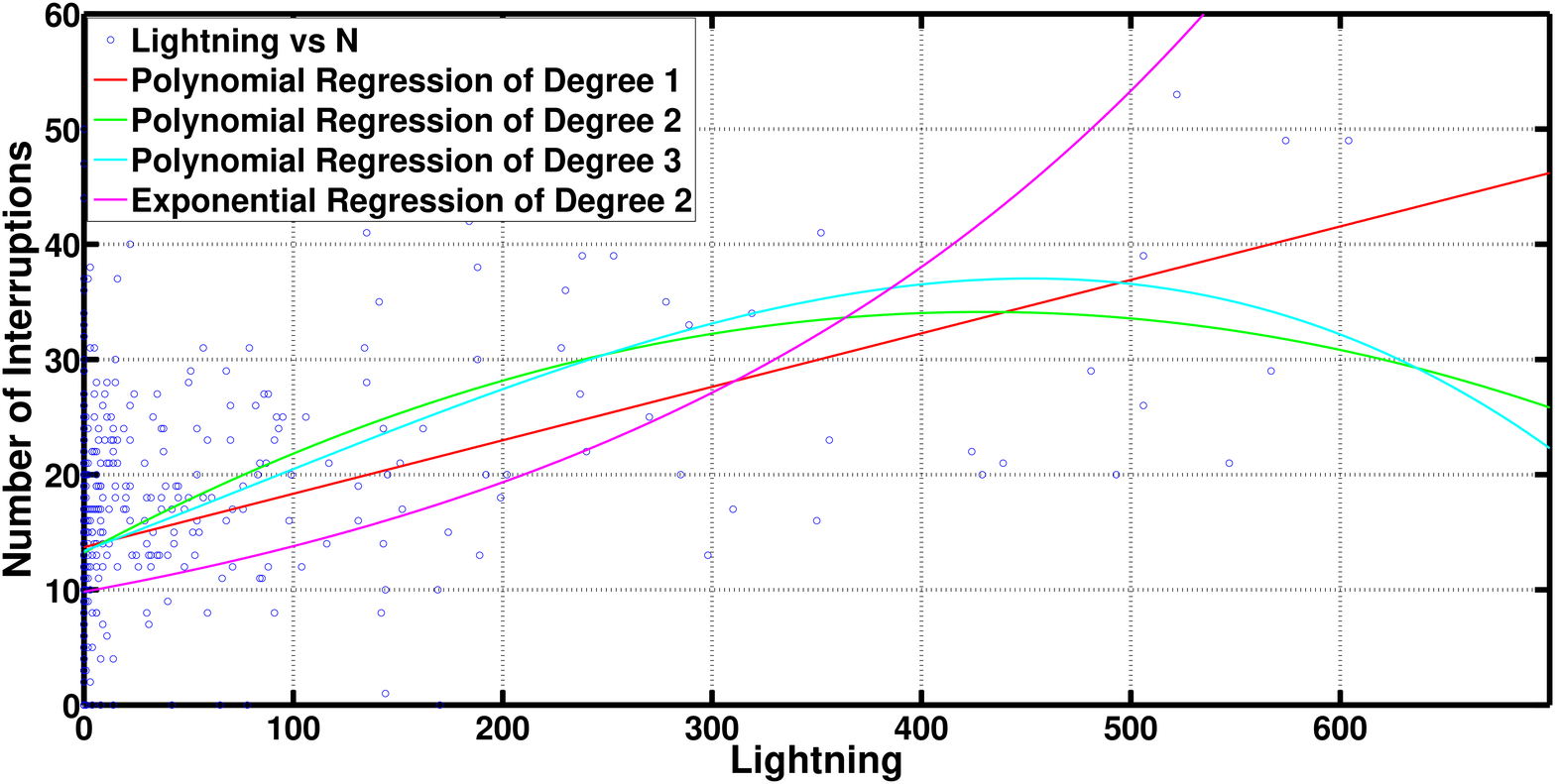}}
\caption{(a) Lightning $L$ (b) The polynomial and exponential regression for analyzing the number of $N_{L}$ respond to lightning $L$}
\label{fig:Light}
\end{figure*}

\subsection{Air pressure}

Air pressure is the pressure within the atmosphere of Earth, which is highly connected with other weather parameters, such as raining, heat storm, and wind speed. The variation of air pressure $A$ for one utility management area is plotted in Fig. \ref{fig:Air}(a). The relationship between air pressure $A$ and $N_{A}$ is analyzed by both polynomial and exponential regression models as shown in Fig. \ref{fig:Air}(b). Polynomial regression with $3$-th degree has a better performance for $A$ with less SSE and RMSE, and greater R-square according to Table \ref{SSE}. Therefore, the relationship function between $N_{A}$ and $A$ can be expressed as: $N_A=\beta^{A}_0+\beta^{A}_1{A}+\beta^{A}_2{{A}^2}+\beta^{A}_3{{A}^3}$.

\subsection{Lightning}
Highly-scaled electrical discharges between the cloud and a piece of earth are called lightning strikes $L$. This natural phenomenon may strike the phase conductors, the tower or shield wires causing backflash, and a piece of nearby ground generating transient overvoltage. The energy of the lightning flash may exceed the thermal limit of the struck object causing thermal failure \cite{L1}. In addition, the combined effects of strong winds and rain are generally accompanied with $L$, Thus, $L$ can have a random, though important, effect on the numbers of $N_L$ in distribution networks. Based on the goodness-of-fit performance results in Table \ref{SSE}, for lightning $L$, polynomial regression with $3$-th degree has a better performance with less SSE and RMSE, and greater R-square. Therefore, the relationship function between $N_L$ and $L$ can be determined by: $N_{L}=\beta^{L}_0+\beta^{L}_1{L}+\beta^{L}_2{L^2}+\beta^{L}_3{{L}^3}$.
Fig. \ref{fig:Light}(a) plots the numbers of $L$ in one utility management area, and both polynomial and exponential regression models are adopted for analysis of the effect of $L$ on $N_L$ as shown in Fig. \ref{fig:Light}(b).

\section{Power interruption forecasting framework}
In this section, taking the polynomial and exponential regression models derived for common weather parameters as inputs, a MLP based forecasting framework is developed for forecasting the daily numbers of $N$ and $M$ in smart grid distribution networks. Additionally, a modified ELM based algorithm is proposed to train, validate, and test the proposed forecasting framework.

\subsection{MLP based forecasting framework}
The MLP is an artificial neural network composed of input, hidden, and output layers in a feed-forward architecture. The MLP neurons in the input layer receive sample data for analysis, and the neurons of the output layer give the network results out. In addition, a MLP network structure is made up by usually one, but occasionally more than one hidden layers between input and output layers. The hidden layer neurons learn the non-linear relationship between the inputs and outputs. In the MLP, all neurons are implemented with non-linear activation functions and each MLP layer is fully connected to the next layer. The mathematical expression of the outputs of the MLP can be defined as follows:
\begin{equation}
Y=F(b+\sum_{j=1}^{m}v_j[\sum_{i=1}^{n}{G(w_{ij}x_i+b_j)}])
\end{equation}
where $x_i$, $i=1,..,n$, is the input value; $Y$ is the output value; $w_{ij}$, $j=1,...,m$, is the weight of connection between the $i$th input neuron and $j$th hidden neuron; $v_j$ is the weight of connection between the $j$th hidden neuron and output neuron; $b$ and $b_j$ are the bias values of the corresponding output neuron and $j$th hidden neuron; and $F(\cdot)$ and $G(\cdot)$ are the activation functions of output and hidden neurons, respectively.

\begin{figure}[tbp]
\centering
  \includegraphics[width=8cm]{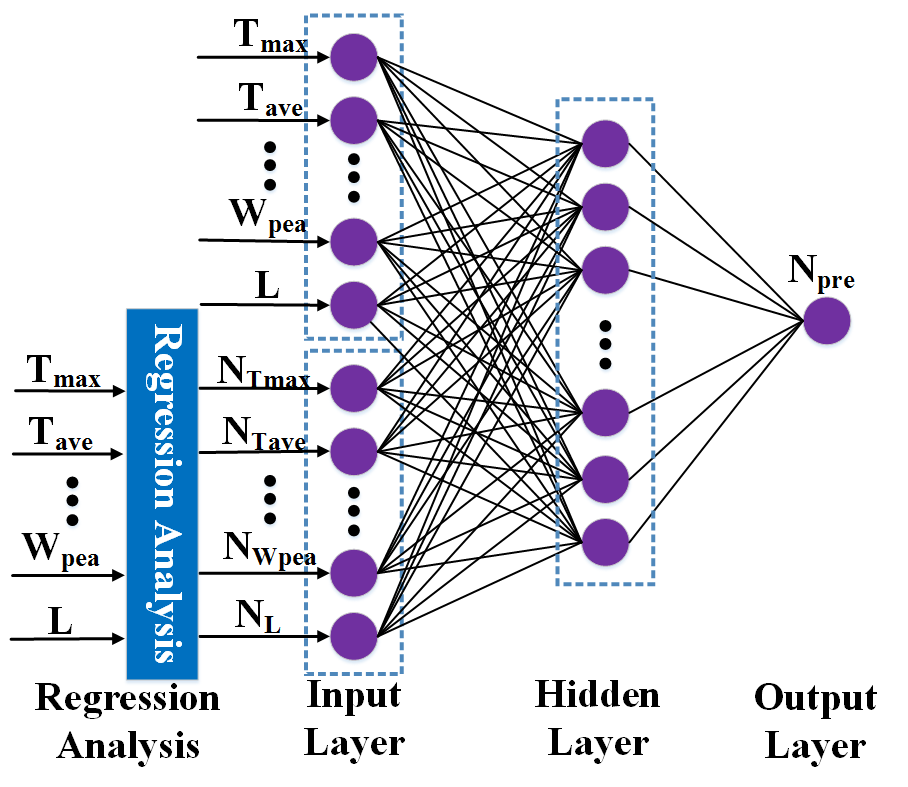}
\caption{The proposed MLP based forecasting framework}
\label{fig:MLP}       
\end{figure}

The MLP can be applied to analyze the combined effect of various weather parameters on the reliability performance of smart grid distribution networks. However, compared with regression models, traditional MLPs usually have difficulty in handling with a large number of input variables due to the time needed for variable preprocessing and the possibility of model overfitting \cite{Hybrid1}. In this paper, a hybrid model integrating MLPs and parametric regression models is proposed to analyze the combined effect of common weather parameters on the numbers of $N$ and $M$. The input layer of the proposed MLP based forecasting framework contains 24 neurons including $T_{\textrm{max}}$, $T_{\textrm{ave}}$, $T_{\textrm{min}}$, $H$, $C$, $W_{\textrm{pea}}$, $W_{\textrm{ave}}$, $W_{\textrm{sus}}$, $P_{\textrm{rain}}$, $A$, and $L$, and corresponding daily numbers of $N$ and $M$ derived by their regression models. The output layer is for the forecasted daily numbers of $N$ and $M$. To restrict the net capacity, one hidden layer including ten neurons is included in the MLP. The network structure of the MLP based forecasting framework is shown in Fig. {\ref{fig:MLP}}.

\subsection{Extreme learning machine (ELM) algorithm}
Acting as a feed-forward neural network learning system, ELM is recently becoming an emerging and efficient algorithm to estimate the weighting parameters of the MLP structure \cite{ELM1}, where the model training is transformed into a matrix calculation problem. Assume the activation function $F(\cdot)$ at the hidden layer to be a constant function, the MLP model then can be represented in the matrix form:
\begin{equation}
\label{ELM1}
\boldsymbol{Y}=\sum_{j=1}^{m}v_j[\sum_{i=1}^{n}{G(w_{ij}x_i+b_j)}]=\boldsymbol{H}\boldsymbol{v}
\end{equation}
where $\boldsymbol{H}\in{\mathbb{R}^{n\times{m}}}$ is the hidden layer output matrix, and $\boldsymbol{v}=[v_1,...,v_m]^T$ represents the output weight vector. Since $w_{ij}$, $i=1,...,n$, and ${b}_j$ are determined randomly for each hidden layer neuron $j=1,..,m$, the objective of ELM algorithm is to calculate $\boldsymbol{v}$ in order to formulate the MLP model. Given a training set $\{\boldsymbol{X},\boldsymbol{Y}\}$, where $\boldsymbol{X}=\{x_1,...,x_n\}$, $\boldsymbol{v}$ can be derived by $\boldsymbol{v}=\boldsymbol{H}^{\dagger}\boldsymbol{Y}$, where $\boldsymbol{H}^{\dagger}$ can be be calculated through orthogonal projection: $\boldsymbol{H}^{\dagger}=(\boldsymbol{H}^T\boldsymbol{H})^{-1}\boldsymbol{H}^T$.

However, ELM algorithm may have a high training error when inappropriate $w_{ij}$ and ${b}_j$ are selected. In order to boost the training performance of algorithm, a self-adjusting parameter $\lambda=\|Y\|^{\delta}$, $\delta\in{[1,2]}$, is introduced into the diagonal elements of $\boldsymbol{H}^T\boldsymbol{H}$: $\boldsymbol{H}^{\dagger}=(\boldsymbol{H}^T\boldsymbol{H}+\lambda)^{-1}\boldsymbol{H}^T$. Therefore, the MLP learning problem is converted into a least square problem defined as:
\begin{equation}
\label{ELM2}
\min_{\boldsymbol{v}}~~\|\boldsymbol{v}\|^{\delta_1}+\lambda\|\boldsymbol{H}\boldsymbol{v}-\boldsymbol{Y}\|^{\delta_2}
\end{equation}
where $\delta_1,\delta_2>0$. The modified ELM algorithm intends to minimize both the training error and the norm of output weights, where $\lambda$ is a parameter to balance the two. The procedures of the modified ELM algorithm are summarized as Table \ref{Step2}. The convergence of the proposed ELM learning algorithm can be proved based on the expansion of \emph{Theorem 3.1} in \cite{ELM2} shown as follows:
\begin{theorem}\label{th:convergence}
Assume $F(\boldsymbol{v})=\boldsymbol{H}\boldsymbol{v}-\boldsymbol{Y}$ is semismooth over $L(\boldsymbol{v}^{0})$ and the level set: $L(\boldsymbol{v}^{0})=\{\boldsymbol{v}\in{\mathbb{R}^{m}}:f(\boldsymbol{v})\leq{f(\boldsymbol{v}^{0})}\}$, where $f(\boldsymbol{v})=\frac{1}{2}\|F(\boldsymbol{v})\|^2$, be bounded. Let $\{\boldsymbol{v}^{k}\}$ be the series generated by the modified ELM algorithm, then the algorithm terminates in finite iterations or satisfies $lim_{k \to \infty} \|F(\boldsymbol{v}^{k})\|=0$.
\end{theorem}

\begin{table}[h!]
  \centering
  \caption{
  \label{Step2}
    \vspace*{-0.0em}Formulated MLP based forecasting framework}
    \begin{tabular}{p{8cm}}
      \toprule
Parametric regression analysis and modified ELM algorithm \vspace*{.3em}\\ \hline
\textbf{Phase 1 - Data collection \& parametric regression analysis:}   \vspace*{.2em}\\
\hspace*{1em}a) Collect the daily numbers of $N$ and $M$ in one utility area; \vspace*{.2em}\\
\hspace*{1em}b) Collect the common weather parameters including $T_{\textrm{max}}$, $T_{\textrm{ave}}$, \vspace*{.2em}\\
\hspace*{2em}$T_{\textrm{min}}$, $H$, $C$, $W_{\textrm{pea}}$, $W_{\textrm{ave}}$, $W_{\textrm{sus}}$, $P_{\textrm{rain}}$, $A$, and $L$ for the area; \vspace*{.2em}\\
\hspace*{2em}\textbf{for} each common weather parameter \textbf{do} \vspace*{.2em}\\
\hspace*{1em}1: Implement $f(\boldsymbol{x},\boldsymbol{\beta}^\textrm{pol})=\beta^\textrm{pol}_0+\beta^\textrm{pol}_1{\boldsymbol{x}}+\beta^\textrm{pol}_2{\boldsymbol{x}^2}+\cdot\cdot\cdot+\beta^\textrm{pol}_n{\boldsymbol{x}^n}$, \vspace*{.2em}\\
\hspace*{2em} $n=1,2,3$, for polynomial regression analysis; \vspace*{.2em}\\
\hspace*{1em}2: Implement $f(\boldsymbol{x},\boldsymbol{\beta}^\textrm{ex})=\beta^\textrm{ex}_0+\beta^\textrm{ex}_1\exp(\beta^\textrm{ex}_2{\boldsymbol{x}})+\beta^\textrm{ex}_3\exp(\beta^\textrm{ex}_4{\boldsymbol{x}})$, \vspace*{.2em}\\
\hspace*{2em} for two-term exponential regression analysis; \vspace*{.2em}\\
\hspace*{1em}3: Analyze the goodness-of-fit of derived regression models; \vspace*{.2em}\\
\hspace*{1em}4: Derive the predicted number of $N$ and $M$ by the optimal \vspace*{.2em}\\
\hspace*{2em}regression model. \vspace*{.2em}\\
\hspace*{2em}\textbf{end for} \vspace*{.2em} \\ \hline
\textbf{Phase 2 - ELM based learning algorithm:}   \vspace*{.2em}\\
\hspace*{1em}a) For each hidden neuron $j=1,..,m$, randomly determine its \vspace*{.2em}\\
\hspace*{2em}input layer weights $w_{ij}$, $i=1,...,n$, and the bias value ${b}_j$; \vspace*{.2em}\\
\hspace*{1em}b) Calculate the hidden layer output matrix $\boldsymbol{H}$ using both  \vspace*{.2em}\\
\hspace*{2em} common weather parameter data and corresponding the \vspace*{.2em}\\
\hspace*{2em} number of $N$ and $M$ derived by Phase 1; \vspace*{.2em} \\
\hspace*{1em}c) Define the self-adjusting parameter $\lambda=\|Y\|^{\delta}$, $\delta\in{[1,2]}$;\vspace*{.2em} \\
\hspace*{1em}d) Obtain the output weight vector $\boldsymbol{v}$ by solving the problem: \vspace*{.2em} \\
\hspace*{2em}$\boldsymbol{v}=\boldsymbol{H}^{\dagger}\boldsymbol{Y}$, where $\boldsymbol{H}^{\dagger}=(\boldsymbol{H}^T\boldsymbol{H}+\lambda)^{-1}\boldsymbol{H}^T$.\vspace*{.2em} \\
\hspace*{2em}Or equivalently, $\min_{\boldsymbol{v}}~~\|\boldsymbol{v}\|^{\delta_1}+\lambda\|\boldsymbol{H}\boldsymbol{v}-\boldsymbol{Y}\|^{\delta_2}$.\vspace*{.2em} \\
\toprule
\end{tabular}
\end{table}

\section{Evaluation of proposed interruption prediction method}

\begin{figure}[h]
\centering
  \includegraphics[width=8cm]{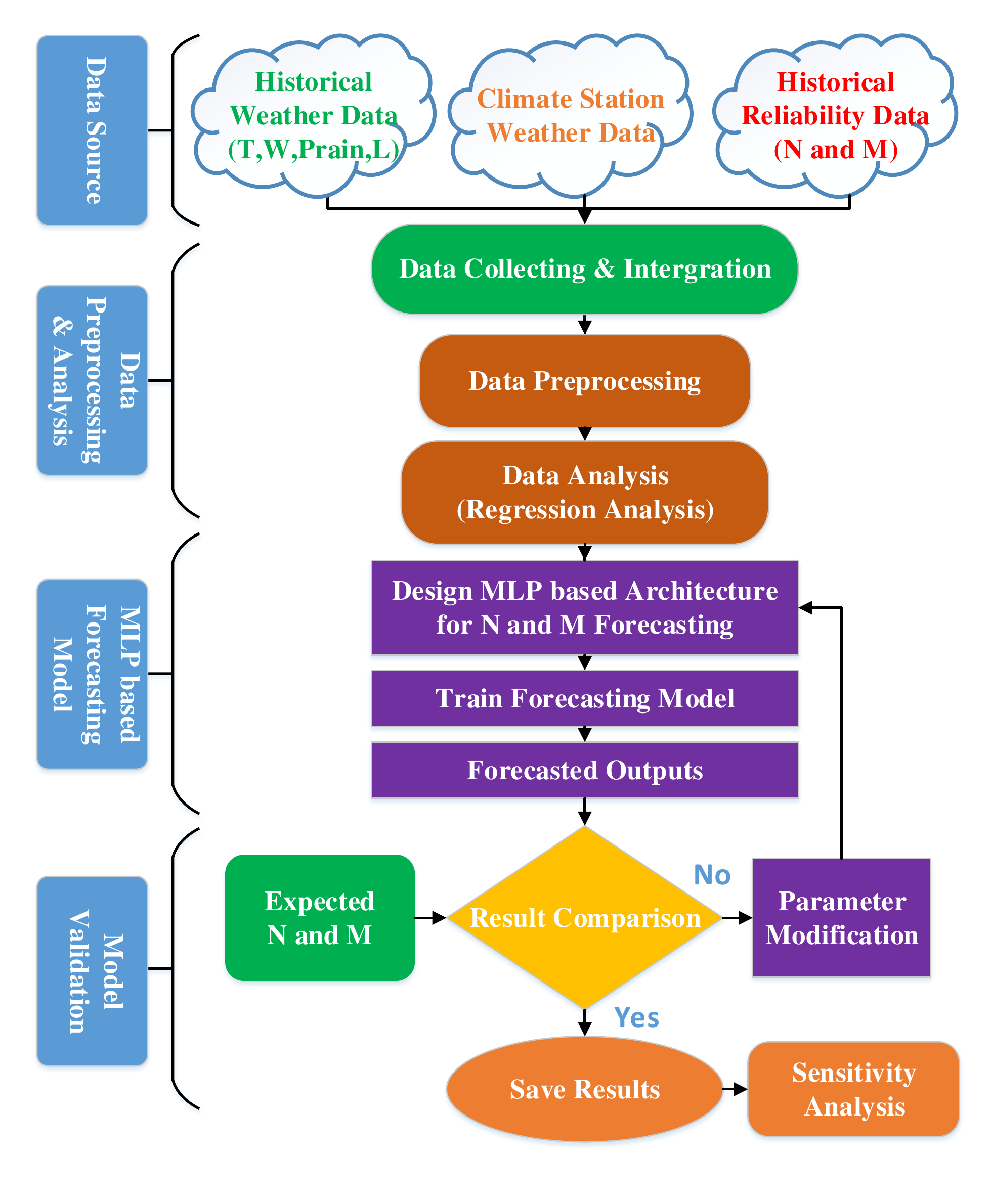}
\caption{The flowchart of the formulated MLP based forecasting model for power distribution networks}
\label{fig:Flow}       
\end{figure}

The implementation of the proposed forecasting framework contains four main parts including: 1) data collection \& preprocessing; 2) parametric regression analysis; 3) MLP based model formulation and training; and 4) sensitivity analysis, in which the flowchart is detailed explained in Fig. \ref{fig:Flow}.

1) \emph{Data Collection \& Preprocessing:} The sustained and momentary power interruption data is collected from one utility management area ranging from Jan. 1st 2015 to Apr. 30th 2017. The interval data is recorded daily, annotated with timestamp. Additionally, one climate station, which is nearest to the central point of this management area, is selected for collecting hourly common weather data including
$T_{\textrm{max}}$, $T_{\textrm{ave}}$, $T_{\textrm{min}}$, $H$, $C$, $W_{\textrm{pea}}$, $W_{\textrm{ave}}$, $W_{\textrm{sus}}$, $P_{\textrm{rain}}$, and $A$. The weather data is noted hourly ranging from Jan. 1st 2015 to Apr. 30th 2017. Correspondingly, the hourly lightning strike $L$ data is collected from the control center of the electric utility located inside this management area.

2) \emph{Parametric Regression Analysis:}
Both polynomial regression with the $n$-th degree, $n=1,2,3$, and two-term exponential regression are implemented for the analysis of the numbers of $N$ and $M$ in the management area response to various weather parameters. Based on the goodness-of-fit results in Table \ref{SSE}, for $T_{\textrm{max}}$, $T_{\textrm{ave}}$, $T_{\textrm{min}}$, $W_{\textrm{sus}}$, $A$, and $L$, polynomial regression with $3$-th degree has a better performance with less SSE and RMSE, and greater R-square. Exponential regression with $2$ terms has a better performance for $H$, $W_{\textrm{pea}}$, $W_{\textrm{ave}}$, and $P_{\textrm{rain}}$, while, for $C$, polynomial regression with $2$-rd and $3$-th degrees and two term exponential regression have similar performances in fitting. Based on these regression models, the daily numbers of $N$ and $M$ can be derived by model fitting.

3) \emph{MLP based Forecasting Framework:}
The input layer of the formulated MLP network contains all common weather parameters and corresponding daily numbers of $N$ and $M$ derived by their regression models for 24 neurons. The hidden layer is set to be one with ten neurons, and the output layer is for the target numbers of $N$ and $M$. The proposed modified ELM algorithm is used for training, validating, and testing the formulated MLP network, in which 60\% of the collected dataset (Jan. 1st 2015 to May 20th 2016) is for the network training, 15\% of the dataset (May 20th 2016 to Oct. 20 2016) is for the network validation, and the remaining 25\% of the dataset (Oct. 20 2016 to Apr. 30 2017) is for the network testing.

\begin{figure} [tbp] \centering
\subfigure[Daily number of $N$] {\label{fig:a}\includegraphics[width=8cm]{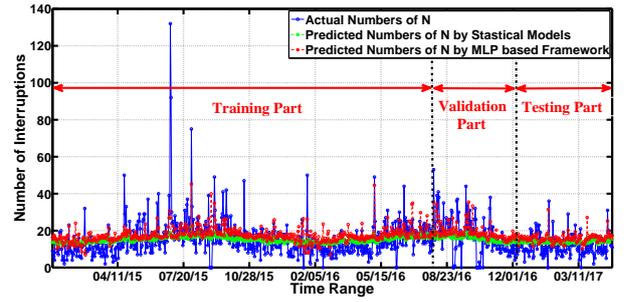}}

\subfigure[Daily number of $M$] { \label{fig:b}\includegraphics[width=8cm]{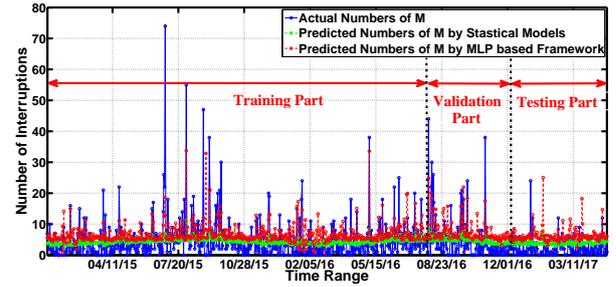}}
\caption{ For one utility management area, (a) the forecasted number of $N$ derived by the MLP neural network and the actual number of $N$ (b) the forecasted number of $M$ derived by the MLP neural network with the actual number of $M$ ranging from Jan. 1st 2015 to Apr. 30th 2017}
\label{fig:Neural}
\end{figure}

\begin{figure}[tbp]
\centering
  \includegraphics[width=8cm]{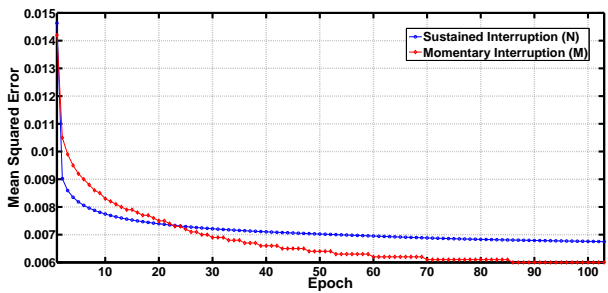}
\caption{Training error of the modified ELM algorithm for forecasting the daily number of $N$ and $M$}
\label{fig:MSE}       
\end{figure}

For all datasets in training, validating, and testing parts, Fig. \ref{fig:Neural}(a) presents the actual numbers of $N$ and the predicted numbers of $N$ derived by statistical models \cite{Arif2} and the proposed MLP based framework. In this figure, we can find that, compared with statistical models, the proposed MLP based framework derives better predicted numbers of $N$. In particular, the proposed MLP based framework yields a Mean-Squared Error (MSE) of $315.4$, which is a $8.77\%$ reduction relative to statistical models. Correspondingly, Fig. \ref{fig:Neural} (b) compares the predicted numbers of $M$ derived by statistical models and the proposed MLP based framework with the actual numbers of $M$, respectively. We can also see that the proposed MLP based framework achieves better predicted numbers of $M$ response to statistical models, in which the proposed MLP based framework achieves a $31.3$ MSE and $61.37\%$ less than the results of statistical models. Furthermore, the training performance and convergence of the modified ELM algorithm for the MLP based framework are shown in Fig. \ref{fig:MSE}, where the y-axis describes the variation of MSE for each training epoch.

\begin{figure}[tbp]
\centering
\includegraphics[width=8cm]{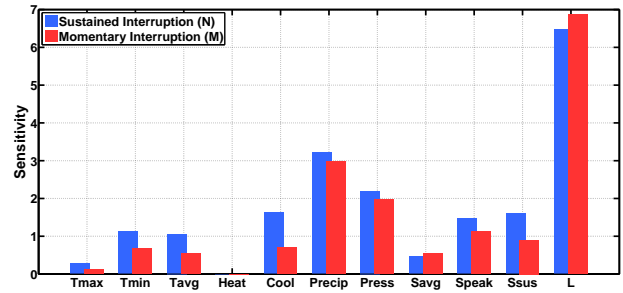}
\caption{Sensitivity analysis of $N$ and $M$ with each weather parameter}
\label{fig:Sens}       
\end{figure}

4) \emph{Sensitivity Analysis:}
The sensitivity of the output to various input perturbations is an important issue in the design and implementation of the MLP based forecasting framework. Therefore, the sensitivity analysis is implemented to analyze the impact of each weather parameter on the daily numbers of $N$ and $M$, respectively. The sensitivity is calculated by the first-order derivative of system network function with respect to the system parameters, which denotes the degree of influence of parameter variations on the network function. Fig. \ref{fig:Sens} presents the sensitivity of each weather parameter response to the daily numbers of $N$ and $M$, respectively.
In this figure, we can find that lightning strike $L$ is the most important weather parameter that has an influence on the daily numbers of $N$ and $M$, while heat degree days $H$ has the least impact on the numbers of $N$ and $M$. This phenomenon can be explained that the most numbers of $N$ and $M$ happen ranging from June to September during one year, which is the raining season for the Florida and lightning strikes happen most frequently. Since the temperature of Florida almost keeps above $65{^{o}F}$, heat degree days have small values during one year.

\section{Conclusion}
This paper presents a MLP based framework to forecast the daily numbers of sustained and momentary interruptions in smart grid distribution networks using time series of common weather data. A modified ELM based learning algorithm is proposed to train, validate, and test the proposed framework, whose convergence is proved. Essentially, compared with traditional statistical models, the proposed framework can reduce MSE by $8.77\%$ and $61.37\%$ for sustained and momentary interruption forecasting, respectively. In addition, we can derive the sensitivity of each common weather parameter with respective to the daily numbers of power interruptions. For the utility management area in Florida, we can find that the lightning strike is the most important common weather parameter impacting on the reliability performance of the smart grid distribution networks, while the heat degree days have the least impacts. In the future, the other factors like power system equipment failure rates and aging of distribution network components can also be integrated as inputs for the proposed framework.

%
%



\end{document}